\documentclass[%
 reprint,
 amsmath,amssymb,
 aps,
 prx,
 floatfix,
 superscriptaddress,
]{revtex4-2}

\usepackage{graphicx}      
\usepackage{dcolumn}       
\usepackage{bm}            
\usepackage{xcolor}        

\usepackage{mathrsfs}
\usepackage{algorithm}
\usepackage{algpseudocode}
\usepackage{chngcntr} 
\usepackage{hyperref}
\hypersetup{
    colorlinks=true,       
    linkcolor=blue,        
    citecolor=blue,        
    filecolor=magenta,     
    urlcolor=blue          
}

\begin{document}

\title[Article Title]{Collider-quality electron bunches from an all-optical plasma photoinjector}

\author{Arohi Jain}
\email{arohi.jain@stonybrook.edu}
\affiliation {Department of Physics and Astronomy, Stony Brook University, Stony Brook, USA}
\author{Jiayang Yan}
\affiliation {Department of Physics and Astronomy, Stony Brook University, Stony Brook, USA}
\author{Jacob R. Pierce}
\affiliation{ Department of Electrical and Computer Engineering, University of California, Los Angeles, California 90095, USA}

\author{Tanner T. Simpson}
\affiliation{University of Rochester, Laboratory for Laser Energetics, Rochester, New York 14623, USA}

\author{Mikhail Polyanskiy}
\affiliation{Accelerator Test Facility, Brookhaven National Laboratory, Upton, NY 11973, USA}
\author{William Li}
\affiliation{Accelerator Test Facility, Brookhaven National Laboratory, Upton, NY 11973, USA}
\author{Marcus Babzien}
\affiliation{Accelerator Test Facility, Brookhaven National Laboratory, Upton, NY 11973, USA}
\author{Mark Palmer}
\affiliation{Accelerator Test Facility, Brookhaven National Laboratory, Upton, NY 11973, USA}

\author{Michael Downer}
\affiliation{Department of Physics, University of Texas at Austin, Austin TX 78712, USA}
\author{Roman Samulyak}
\affiliation{Department of Physics and Astronomy, Stony Brook University, Stony Brook, USA}
\author{Chan Joshi}
\affiliation{ Department of Electrical and Computer Engineering, University of California, Los Angeles, California 90095, USA}
\author{Warren B. Mori}
\affiliation{ Department of Electrical and Computer Engineering, University of California, Los Angeles, California 90095, USA}

\author{John P. Palastro}
\affiliation{University of Rochester, Laboratory for Laser Energetics, Rochester, New York 14623, USA}

\author{Navid Vafaei-Najafabadi}
\email{navid.vafaei-najafabadi@stonybrook.edu}
\affiliation {Department of Physics and Astronomy, Stony Brook University, Stony Brook, USA}

\begin{abstract}
We present a novel approach for generating collider-quality electron bunches using a plasma photoinjector. The approach leverages recently developed techniques for the spatiotemporal control of laser pulses to produce a moving ionization front in a nonlinear plasma wave. The moving ionization front generates an electron bunch with a current profile that balances the longitudinal electric field of an electron beam-driven plasma wave, creating a uniform accelerating field across the bunch. Particle-in-cell (PIC) simulations of the ionization stage show the formation of an electron bunch with 220 pC charge and low emittance ($\varepsilon_x = 171$ nm rad, $\varepsilon_y = 76$ nm rad). Quasistatic PIC simulations of the acceleration stage show that the bunch is efficiently accelerated to 24 GeV over 2-meters with a final energy spread of less than 1\% and emittances of $\varepsilon_x = 189$ nm rad and $\varepsilon_y = 80$ nm rad. This high-quality electron bunch meets the requirements outlined by the Snowmass process for intermediate-energy colliders and compares favorably to the beam quality of proposed and existing accelerator facilities. The results establish the feasibility of plasma photoinjectors for future collider applications making a significant step towards the realization of high-luminosity, compact accelerators for particle physics research.
\end{abstract}

\maketitle

\section{INTRODUCTION}
Two decades ago, the so-called ``dream-beam'' papers reported the first demonstration of monoenergetic relativistic electron bunches from the interaction of ultrashort laser pulses with plasma \cite{Mangles04,Geddes04,Faure04}. In the years since, the plasma accelerator community has made great strides towards generating bunches with potential application to a collider, which would fulfill the original dream of replacing the MV/m fields of radio frequency cavities with the GV/m fields of nonlinear plasma waves. As part of the Snowmass process, where the high energy physics community outlines their priorities for the field, the plasma accelerator community laid out a vision for plasma-based colliders at TeV-scale energies \cite{benedetti2022linear} as well as an intermediate demonstration facility at $\sim$10-50 GeV\cite{benedetti2022whitepaper}. Realizing the vision of a plasma-based collider requires electron bunches with hundreds of pC of charge at a normalized emittance below 100 nanometers \cite{benedetti2022linear, benedetti2022whitepaper}. In addition, the energy spread must be less than 1\%  for compatibility with the final focusing section of a particle collider and to maximize the cross-section of collisions. While the plasma accelerator community has explored innovative ideas for producing such bunches \cite{faure2017plasma}, simultaneous achievement of all three requirements---high charge, low emittance, and low energy spread---has proved elusive \cite{fuchs2022snowmass}. 

\begin{figure}[!t]
    \centering
    \includegraphics[width=0.5\textwidth]{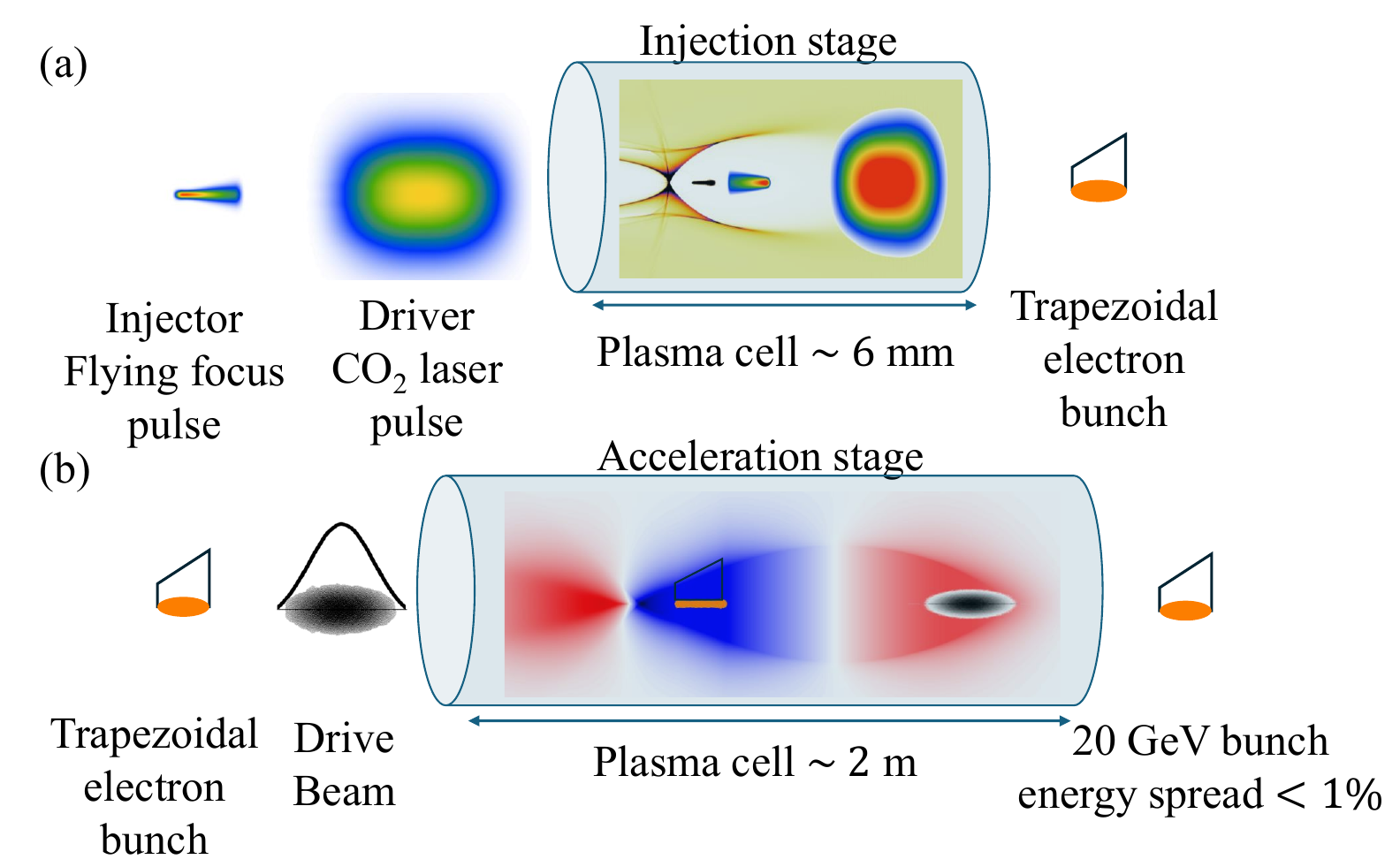}
    \caption{Production of collider-quality electron beams with a flying focus photoinjector. (a) A long-wave infrared laser pulse partially ionizes a gas and drives a nonlinear plasma wave. A trailing, shorter-wavelength flying focus pulse further ionizes the gas within the plasma wave, creating a moving ionization front. The electrons freed by the moving ionization front form a bunch with a trapezoidal current profile. (b) The trapezoidal bunch is further accelerated in a plasma wakefield accelerator. The trapezoidal shape of the bunch flattens the accelerating field of the plasma wave, ensuring uniform acceleration across the bunch. For the simulated parameters (Tables  \ref{tab:Parameters} and \ref{tab:acc_stage_Parameters}), the bunch is accelerated to 20 GeV over 2-meters and has an energy spread of less than 1\%, which meets the quality requirements for a collider.}
    \label{fig:Schematic}
\end{figure}

In this paper, we demonstrate how emerging techniques for controlling the spatiotemporal properties of a laser pulse enable an all-optical ``plasma photoinjector" that generates an electron bunch satisfying all requirements for plasma collider applications. The approach, illustrated in Fig. \ref{fig:Schematic}, is based on applying the flying focus scheme \cite{froula2019flying} to the two-color ionization injection method \cite{xu2014low,Schroeder2014,schroeder2015ultra} in a laser wakefield accelerator (LWFA). Earlier applications of the flying focus to LWFA aimed to extend the LWFA interaction without the need for external guiding structures \cite{Palastro2020dephasing}. Here, we present the first use of the flying focus for injection. The consequence is a dramatic enhancement in injected-beam luminosity---by more than an order of magnitude---over previous approaches. While this method is described in the context of a beam with 200 pC of charge to satisfy collider requirements \cite{benedetti2022whitepaper}, the framework developed in this paper for systematically adjusting the beam charge, emittance, and profile enables the production of electron bunches with charges exceeding 1 nC or emittances as low as 10 nm. This makes the approach versatile, with the potential to impact a wide range of applications in advanced accelerator research, from particle colliders to compact light sources.

The flying focus injector pulse naturally produces a trapezoidal current profile (Fig. \ref{fig:Schematic}). This distinctive shape, first proposed by M. Tzoufras \textit{et al}. in 2008 \cite{Tzoufras2008PRL}, balances the accelerating field across the entire bunch ensuring low energy spreads. As a result, the flying focus injector provides the first technique capable of directly shaping the bunch into this optimal form, making making collider-quality beam generation feasible. 
To illustrate this, the bunch generated by the photoinjector is accelerated in a plasma wakefield accelerator to 24 GeV with sub-percent energy spread that does not grow due to the unique shape of the current profile. The concurrent realization of high charge, low emittance, and low energy spread, demonstrated in this work, represents a major advance toward the realization of a plasma-based collider.

In the next section, we introduce the two-color ionization injection scheme, which forms the foundation of this work. Section III presents the underlying physics that enable control of the injected beam properties using a flying-focus laser. Finally, Section IV discusses the resulting beam quality, demonstrating collider-relevant performance with hundreds of picocoulombs of charge, sub-200 nm emittance, and sub-percent energy spread at energies reaching tens of GeV.

\section{Two-color ionization injection}
In two color ionization injection, a long-wavelength laser pulse serves as the ``driver" of a plasma wave, and a second, shorter-wavelength laser pulse acts as the ``injector." An  injector pulse focused by an ideal lens, as was used in previous work \cite{xu2014low,Schroeder2014,schroeder2015ultra,li2019laser}, co-propagates with the plasma wave, but remains a fixed distance behind it, which constrains properties of the injected bunch. Here, a ``flying focus" injector pulse is used instead. The flying focus pulse features a dynamic focal point, which allows the peak intensity to move with respect to the plasma wave.

\begin{figure*}[!bt]
\centering
\includegraphics[width=.8\textwidth]{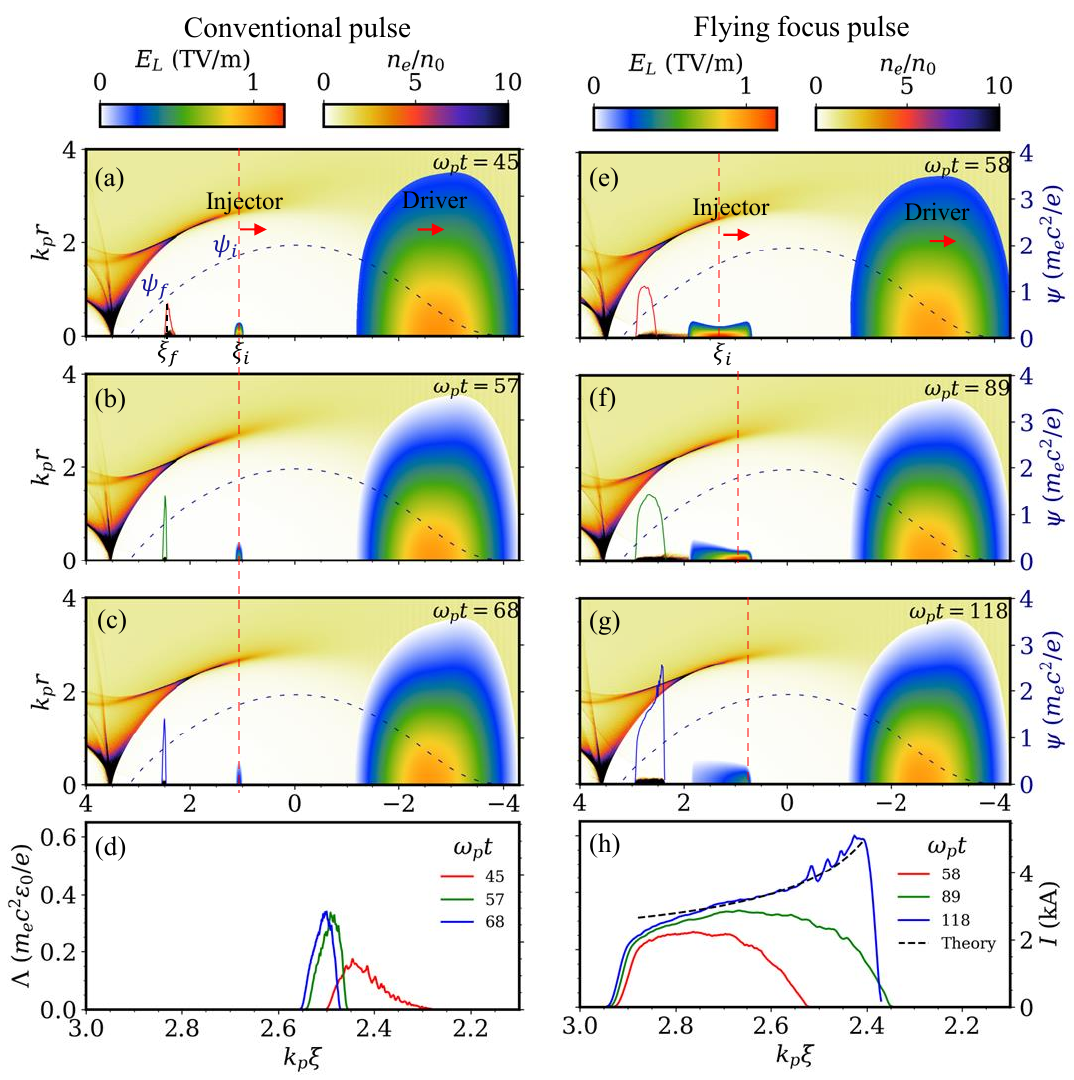}
\vfil
\caption{Comparison of ionization injection using a conventional or flying focus laser pulse. A CO$_2$ laser pulse (driver) partially ionizes the background gas and drives a nonlinear plasma wave (a-c and e-g). An optical laser pulse (injector) trailing the CO$_2$ pulse further ionizes the gas within the plasma wave. The electrons freed by the injector are accelerated longitudinally as they move backward in the moving frame coordinate $\xi = ct -z$ from a higher potential $\psi_i$ to a lower potential $\psi_f$. At which point, they become trapped in the wave. The accumulation of these electrons over time $t$ forms a bunch with a charge per unit length $\Lambda$ (d and h). The conventional laser pulse (a-d) frees electrons at a fixed $\xi$ (red dashed line), leading to a narrowly peaked, triangular current profile (d). The flying focus pulse (e-h) creates a moving ionization front (red dashed line), leading to a trapezoidal current profile (h) that is ideal for beam loading in a subsequent acceleration stage. The contours illustrate the electric field amplitudes of the pulses $E_L$ and electron density $n_e$. The conventional and flying focus pulses have the same peak intensity, focused spot size, and transform-limited pulse duration. See Table \ref{tab:Parameters} for the parameters.}
\label{fig:FF vs conv}
\end{figure*}

Figure \ref{fig:FF vs conv} illustrates the underlying physics of two-color ionization injection and contrasts a typical conventional lens configuration with the novel flying focus configuration introduced here (see end matter for simulation details). Both configurations use the same CO$_2$ laser pulse as the driver  (see Table \ref{tab:Parameters} for parameters). The CO$_2$ pulse propagates through and ionizes a gas, freeing the outer shell electrons and forming a plasma. The ponderomotive force of the pulse then pushes the plasma electrons away from the propagation axis. The ions, which remain stationary over the timescale of the interaction, pull the displaced electrons back towards the axis, creating a region of positive charge density surrounded by a sheath of electrons. The accelerating field within this nonlinear plasma wave, commonly referred to as the plasma bubble \cite{lu2006nonlinear}, exceeds 10 GV/m, which is hundreds of times larger than the conventional accelerators currently in operation. 

A long-wave infrared laser pulse such as a CO$_2$ pulse with a wavelength $\lambda = 9.2 ~\mu m$ is an ideal driver for two-color ionization injection. This is because the magnitude of the accelerating field increases with the normalized vector potential of the pulse, $a_0=0.86\times 10^{-9} \sqrt{I[\mathrm{W}\mathrm{cm}^{-2}](\lambda [\mu \mathrm{m}])^2 }$, where $I$ is the intensity. A strongly nonlinear plasma wave is expected when $a_0>2$ (The CO$_2$ laser pulse in Fig. \ref{fig:FF vs conv} has an $a_0=2.7$). 
The $I\lambda^2$ scaling allows the CO$_2$ pulse to reach these values of $a_0$ at an intensity that is a hundred times smaller than would be required with a near-infrared pulse, such as produced by typically used Ti:Sapphire lasers. The longer laser wavelength also allows for easily attaining larger focal spot sizes. In the simulations shown in Fig.~\ref{fig:FF vs conv}, the resulting plasma wakefield has a diameter of nearly $200~\mu$m, a scale that significantly relaxes synchronization and alignment requirements between the driver and injector in practical implementation.

The intensity of the CO$_2$ pulse is high enough to free the outer shell electrons of a background gas, e.g., the first eight levels of krypton, producing Kr$^{8+}$ (See Appendix A). To initiate ionization injection, a short-wavelength injector pulse ($\lambda=0.4$ $\rm \mu m$) is focused behind the driver to a peak $a_0$ of 0.17, which is selected so that the intensity exceeds the ionization threshold of Kr$^{8+}$. Note that although the $a_0$ of the injector pulse is much smaller than that of the CO$_2$ pulse, it has a much higher intensity due to its shorter wavelength. The injector pulse frees the outer Kr$^{8+}$ electrons in the accelerating phase of the plasma wave. These electrons are then accelerated to relativistic energies and trapped in the wave.

This configuration, in which a driver partially ionizes a gas and excites a large-amplitude plasma wave while an injector triggers additional ionization within the wave, is fundamental to two-color ionization injection. The driver and injector, however, are fungible---that is, they can be replaced with suitable alternatives. For instance, the CO$_2$ pulse can be substituted for a dense electron beam, provided that the parameters are chosen so that the driver excites a large-amplitude plasma wave without fully ionizing the gas, leaving it to be further ionized by the injector.

\begin{table}[t!]
\caption{Simulation parameters for the injection stage. The laser parameters are motivated by the available facilities and near-term upgrade plans of the Accelerator Test Facility at Brookhaven National Laboratory. The plasma is initialized as Kr$^{8+}$ with an electron density $n_0 = 2\times 10^{16}$  cm$^{-3}$. }
\label{tab:Parameters}
\begin{tabular}{l c c c}
 \hline
 \hline
 Parameter & Value & Units\\
 [1ex]
  \hline
  Driver- CO$_2$ laser pulse & & \\
 \hline
Normalized amplitude $a_0$ & 2.7& \\
Wavelength & 9.2 & $\mu$m\\
Spot size (1/e of field) & 105 & $\mu$m\\
Duration (FWHM) & 250 & fs\\
Energy & 5.3 & J\\
[1ex]
 \hline
Injector pulse  & & \\
 \hline
Normalized amplitude $a_0$ & 0.17 & \\
Wavelength & 0.4 & $\mu$m\\
Spot size (1/e of field) & 8.6 & $\mu$m\\
Transform limited duration (FWHM) & 20 & fs\\
Initial delay between injector and driver & 0.4& ps\\
[1ex]
\underline{Conventional pulse}  & & \\
Energy & 6.2 & mJ\\
[1ex]
\underline{Flying focus pulse}  & & \\
Energy & 36 & mJ\\
Focal range $L$& 3.8& mm\\
Focal velocity $v_F$& 1.01 & \\
\hline
\hline
\end{tabular}
\end{table}

\section{Shaping the injected bunch with a flying focus}

The electron bunch formed by the ionization, acceleration, and trapping depends on the dynamics of the injector. The peak intensity of a conventional pulse and location at which it ionizes move at approximately the group velocity $v_g\sim c$ (inside the nonlinear plasma bubble, where the local electron density is nearly zero, the refractive index approaches unity). As a result, the location of ionization is nearly fixed in the moving frame coordinate $\xi = ct - z$ [red dashed line in Figs. \ref{fig:FF vs conv}(a-c)], which provides almost no flexibility to structure the injected bunch. In contrast, the peak intensity of the flying focus injector travels at a specified velocity $v_F$, producing an ionization front that moves in $\xi$ [red dashed line in Figs. \ref{fig:FF vs conv}(e-g)]. The ability to control the velocity of the moving ionization front provides the flexibility needed to structure the longitudinal profile of the injected bunch.

The flying focus pulse required for such an ionization front can be produced using recently developed techniques for spatiotemporal control of laser pulses. While several methods have been demonstrated and proposed \cite{froula2018spatiotemporal, palastro2018ionization, Palastro2020dephasing, turnbull2018ionization,palastro2021laser,miller2023dephasingless,pigeon2024ultrabroadband, sainte2017controlling, caizergues2020phase, oubrerie2022axiparabola}, all share the property that the focal point, and thus the peak intensity, can be made to move at a velocity $v_F$ that is decoupled from the group velocity $v_g$ over a distance $L$ that is much longer than the Rayleigh length. Crucially, when $v_F \neq c$, the motion of the peak intensity in the coordinate $z$ is accompanied by motion in the moving frame coordinate $\xi$:
\begin{equation}
 \frac{d {\xi}}{d {z}} = \frac{1}{ {v}_F} - 1, \label{eqn: xi velocity of focus}   
\end{equation}
where normalized units have been adopted (see Appendix A).
Thus, the flexibility to specify ${v}_F$ provides control over the motion of the peak intensity and ionization front in $\xi$. 

The motion of the ionization front determines the longitudinal profile of the injected bunch. The profile can be derived from the location of ionization $\xi_i$ using the normalized wake potential, $ {\psi}\equiv \left( \phi - A_z \right)$, where $\phi$ and $\bold{A}$ are the scalar and vector potentials of the plasma wave, respectively. The freed electrons will be trapped in the plasma wave if $ {\psi_f} -  {\psi_i} \approx -1$, where $\psi_i$ is the initial wake potential experienced by an electron when it is freed and $\psi_f$ is the minimum wake potential experienced \cite{oz2007ionization, pak2010injection, mcguffey2010ionization}. Near the propagation axis ($r\approx 0$), ${\psi} = {\psi}_{\mathrm{max}}-\frac{\chi}{4} {\xi^2}$, where $0< \chi\leq1$ depends on the strength of the plasma wave \cite{lu2006nonlinear,Tzoufras2008PRL,Xu2014PRL}. Thus, the trapping condition provides a direct connection between the location of ionization ${\xi_i}$ and the location of injection ${\xi_f}$, where the electron reaches the same velocity as plasma wave: ${\xi_f^2} = 4/\chi +  {\xi_i^2}$ [points marked in Fig. \ref{fig:FF vs conv}(a)]. 

The charge per unit length of the injected bunch is given by 
\begin{align}
 {\Lambda} ( {\xi}_f) =\frac{1}{2 \pi}\frac{d  {Q}}{d  {\xi}_f}&= \frac{1}{2 \pi} \frac{dQ}{dz}\frac{dz}{d\xi_i}\frac{d\xi_i}{d\xi_f}. 
\label{eqn: charge per unit length1}     
\end{align} 
The first term on the right-hand-side is the charge per unit length of electrons freed by the injector pulse along its propagation path: $\frac{dQ}{dz}=-\pi w_i^2 n_i $, where $w_i$ is the spot size at the instant of ionization and $n_i$ is the background ion density. The second term describes the motion of the ionization point in the co-moving coordinate. The third term $\frac{d {\xi_f}}{d {\xi_i}}$ is a compression factor, which describes the spatial compression of the injected electrons as they are trapped. For a conventional injector pulse, the location of ionization moves only slightly due to diffraction and $\frac{d {\xi}_i}{d {z}} \approx 0$. As a result, the injected electrons coalesce into a narrowly peaked, triangular profile [Fig. \ref{fig:FF vs conv}(d)]. For a flying focus injector, $\frac{d {\xi}}{d {z}} =  {v}_F^{-1} - 1$ [Eq. \eqref{eqn: xi velocity of focus}], such that   
\begin{equation}
 {\Lambda} ( {\xi}_f)
= -\frac{1}{2}  {w}_i^2 \frac{ {v}_F}{1- {v}_F} \frac{\xi_f}{(\xi_f^2-4/\chi)^{1/2}},
\label{eqn: charge per unit length2}     
\end{equation} 
which is nearly trapezoidal in $\xi_f$ [dashed line in Fig. \ref{fig:FF vs conv}(h)].

\section{Collider-quality bunch generation}
The formation of a charge per unit length ${\Lambda}( {\xi})$ with trapezoidal profile is critical for achieving a low energy spread in a subsequent plasma or laser wakefield acceleration stage. Such a profile flattens the longitudinal electric field in the region occupied by the bunch \cite{Tzoufras2008PRL} so that all of the electrons gain approximately the same energy. The flying focus injector produces a near-ideal trapezoidal bunch with 220 pC of charge and transverse emittances of $\varepsilon_x = 171$ nm rad and $\varepsilon_y = 76$ nm rad. These values meet the collider requirements laid out in the Snowmass parameter set \cite{benedetti2022whitepaper}. The conventional injector pulse, on the other hand, results in a triangularly shaped bunch with 17 pC of charge and a normalized emittance of 140 nm [Fig. \ref{fig:FF vs conv}(d)]. This triangular profile would be ineffective in flattening the longitudinal field, leading to a suboptimal energy spread in the acceleration stage. Moreover, while the emittance meets the requirements for collider applications, the charge is far too low. Thus, the triangular bunch is inadequate for collider applications. 

The total injected charge is given by $Q_\textrm{tot} = -\pi w_i^2 n_i L_i $, where $L_i$ is the distance over which the injector pulses ionize. For a flying focus, this distance is the focal range $L$, which is independent of the spot size and much greater than the Rayleigh range $Z_R = \pi w_i^2 / \lambda$. For a conventional pulse, $L_i \approx Z_R$. Thus, with the same spot size, a flying focus can inject more charge than a conventional pulse by ionizing over a much longer distance ($L \gg Z_R$). While the charge injected by a conventional pulse can be increased by using a larger spot size (Rayleigh range), this has the detrimental effect of increasing the emittance. As an example, simulations \cite{supplement} indicate that a conventional pulse can achieve 236 pC of charge with a spot size of 15 $\mu$m, but the resulting emittance $\varepsilon_x = 1.3$ $\mu$m rad and $\varepsilon_y = 0.69$ $\mu$m rad is too large to meet the Snowmass requirements.

\begin{figure}
    \centering
    \includegraphics[width=0.48\textwidth]{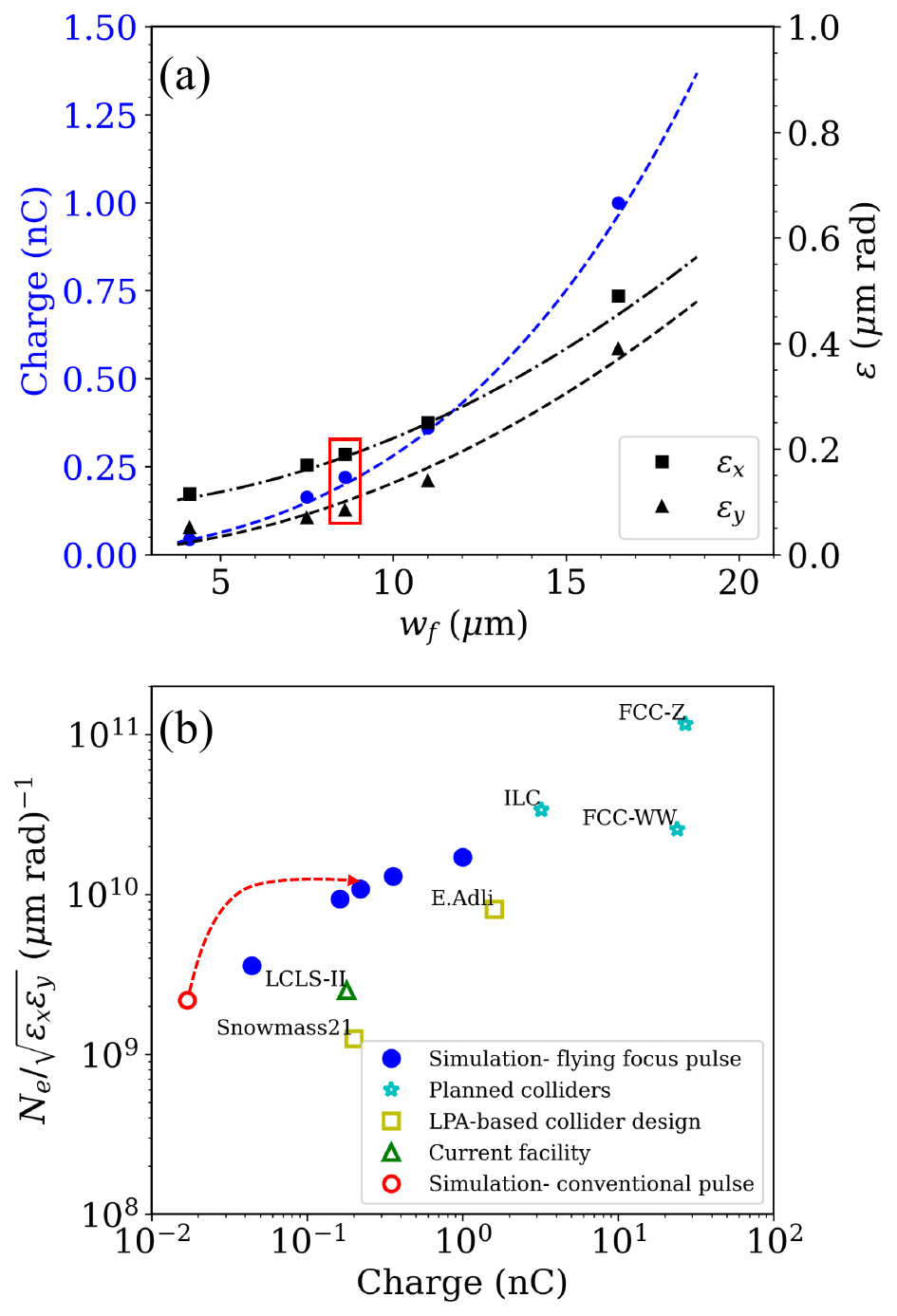}
    \caption{Comparison of the beam quality afforded by the flying focus photoinjector to existing and planned facilities.(a) Increasing the spot size of the flying focus pulse results in more injected charge (blue) at the cost of higher emittances (black and red). The PIC simulalions (points) are in excellent agreement with the theoretical predictions (dashed lines). The red box highlights the electron bunch from Figs. \ref{fig:FF vs conv}(e-h), which has 220 pC of charge, $\varepsilon_x = 171$ nm rad, and $\varepsilon_y = 76$ nm rad and meets the requirements of the Snowmass parameter set \cite{benedetti2022whitepaper}. (b) The electron contribution to the luminosity $N_e / \sqrt{\varepsilon_x \varepsilon_y}$ for the simulation results in (a) alongside planned and existing collider facilities, including E. Adli et al \cite{adli2013beam}, ILC \cite{adli2013beam}, Snowmass \cite{benedetti2022linear}, FCC-WW \& FCC-Z \cite{fcc2019fcc}, and LCLS \cite{raubenheimer2018lcls}. The red dashed arrow indicates the improved luminosity achieved by replacing a conventional pulse with a flying focus pulse for the bunches shown in Fig. \ref{fig:FF vs conv}. }
    \label{fig:comparing facilities}
\end{figure}

The properties of the injected electron bunch can be optimized by adjusting the parameters of the flying focus pulse. For instance, Fig. \ref{fig:comparing facilities}(a) shows that the charge can be scaled to over 1 nC by increasing the spot size of the flying focus. While the increase in charge comes at the cost of higher emittances, each of the simulated spot sizes (points) resulted in an electron bunch with a near-ideal trapezoidal profile (see \cite{supplement} for the results of the 1 nC bunch). The simulation results are in excellent agreement with the theory (dashed lines) for the total charge $Q_\textrm{tot} = -\pi w_i^2 n_i L_i $ and emittance (see Appendix B)\cite{schroeder2014thermal,schroeder2015ultra}. The agreement suggests that the flying focus photoinjector could produce electron bunches with charges exceeding 1 nC or emittances as low as 10 nm. This tunability can be used to generate electron beams with properties that are tailored to a particular application. 

Figure \ref{fig:comparing facilities}(b) demonstrates that the flying focus photoinjector produces bunches with quality comparable to those of proposed conventional colliders and plasma-based colliders. In a collider, the beam quality is measured using the luminosity, i.e., the event rate per unit time per unit area, given by 
$\mathscr{L} = H_D N_1 N_2 f\gamma/(4\pi \sigma_x \sigma_y)$. Here, $N_1$ and $N_2$ are the number of particles in the colliding beams, $f$ is the collision frequency, $\gamma$ is the particle energy, $\sigma_x\sigma_y \propto \sqrt{ \varepsilon_x\varepsilon_y}$ is the cross section of the beam, and $H_D$ is an ${\sim}\mathcal{O}(1)$ geometry-dependent factor. The flying focus injector results in an electron contribution to the luminosity $\mathscr{L} \propto N_e/ \sqrt{\varepsilon_x\varepsilon_y}$ that satisfies the criteria for plasma-based colliders, as outlined by E. Adli et al. \cite{adli2013beam} and the Snowmass report \cite{benedetti2022linear} (blue circles compared to yellow squares). The red dashed arrow highlights the improved luminosity of the flying focus pulse over the conventional pulse for the case presented in Fig. \ref{fig:FF vs conv}. 

\begin{figure*}[!t]
\centering
\includegraphics[width=.95\textwidth]{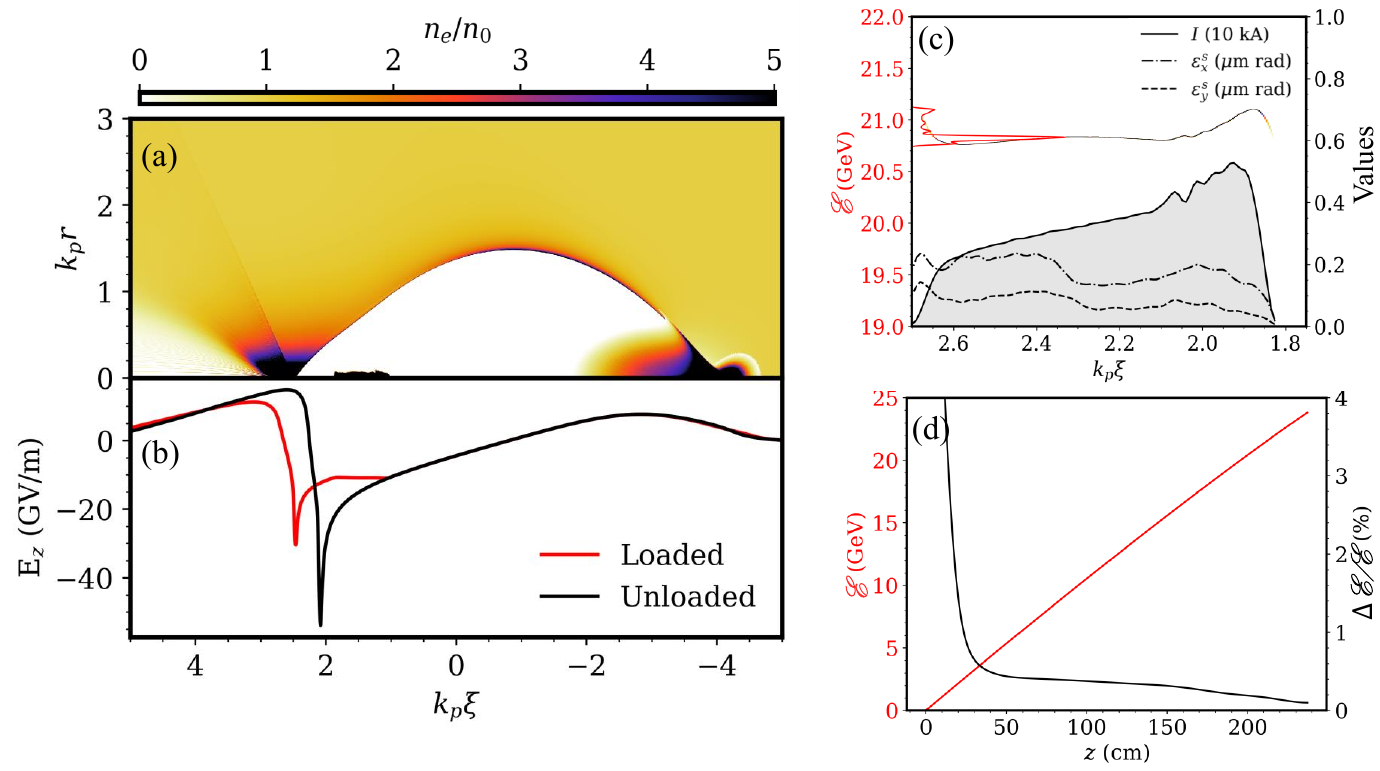}
\vfil
\caption{Acceleration of the trapezoidal electron bunch from the injection stage in an electron-beam driven plasma wave. (a) A drive electron beam expels background electrons which are then pulled back to the axis by the ions, forming an electron density ``bubble''. The trapezoidal electron bunch sits behind the drive beam in the bubble. (b) The longitudinal field of the trapezoidal bunch partially cancels the unloaded field of the bubble (black curve), producing a loaded field that is flattened in the region occupied by the bunch (red curve).
(c) After 2 m of acceleration, the bunch reaches an average energy of 21 GeV (left axis). The trapezoidal profile and low slice emittances are maintained during the acceleration (right axis). (d) The energy gain (left axis) is nearly linear over the length of the accelerator, and the energy spread (right axis) drops precipitously to well below 1\%.}
\label{fig:beam energy}
\end{figure*}

For collider applications, the bunch produced by the flying focus photoinjector must be accelerated to high energy with a narrow energy spread. Achieving a narrow energy spread in a subsequent beam or laser-driven plasma acceleration stage requires placing the bunch in a region of the plasma wave where it can flatten, or ``load,'' the longitudinal electric field $E_z$. This ensures that all electrons within the bunch experience the same longitudinal acceleration. The formalism for calculating the ${\Lambda}(\xi)$ needed to flatten ${E_z}$ was developed in Ref. \cite{Tzoufras2008PRL} for the case of a large plasma bubble $R_b \gg 1$, where $R_b$ is the bubble radius. In this formalism, the injected bunch flattens the field from an initially unloaded profile $E_z = -\tfrac{1}{2}\xi$ to a constant loaded field $E_t$. However, the electron beams available at existing facilities typically drive slightly weaker plasma waves with an $R_b$ between one and three, such that ${E}_z = -\tfrac{1}{2} \chi \xi$ with $0< \chi <1$. In this more general case, the charge per unit length needed to flatten the longitudinal field is given by
\begin{equation}
     {\Lambda}(\xi)  = \frac{1}{4} \sqrt{16\frac{ {E}_t^4}{\chi^4}+ {R}_b^4} - \frac{ {E}_t}{\chi}({ {\xi} -  {\xi}_t}).
    \label{eq: beamload correction}
\end{equation}
This equation describes a trapezoid starting at $\xi=\xi_t$, where the first term is the maximum value and $E_t/\chi$ in the second term is the slope. To ensure that the injected bunch flattens $E_z$ to a desired value of $E_t$, the acceleration stage must be designed so that $R_b$ and $\chi$ produce an approximate equality  between Eqs. \eqref{eqn: charge per unit length2} and \eqref{eq: beamload correction}.

\begin{table}[hbt]
\caption{Simulation parameters for the acceleration stage}
\label{tab:acc_stage_Parameters}
\begin{tabular}{l c c c}
 \hline
 \hline
 Parameter & Value & Units\\
 \hline
 Drive beam electron energy & 20& GeV\\
 Drive beam charge & 0.7 & nC\\
 Drive beam dimensions $\sigma_x$, $\sigma_y$, $\sigma_z$ & 10, 10, 20 & $\mu$m\\
 Driver beam emittances $\varepsilon_x$, $\varepsilon_y$ & 50, 5& $\mu$m rad  \\
 Injected bunch energy& 10& MeV  \\
 Injected bunch charge & 220& pC  \\
 Initial injected bunch emittances $\varepsilon_x$, $\varepsilon_y$ & 171, 76& nm rad  \\
 Background electron density &  $5\times 10^{16}$& cm$^{-3}$ \\
 [1ex]
 \hline
 \hline
\end{tabular}
\end{table}

The electron beam-driven acceleration stage demonstrated here is designed to have a loaded field of approximately $10 \;\mathrm{GV/m}$ for the trapezoidal bunch with 220 pC of charge, $\varepsilon_x = 171$ nm rad, and $\varepsilon_y = 76$ nm rad [Fig. \ref{fig:FF vs conv}(e-h)]. The triangular bunch [Fig. \ref{fig:FF vs conv}(d)] is not considered because (1) it has a much lower charge and (2) its spatial profile does not flatten the accelerating field, which would lead to a higher energy spread. Both of these make the triangular bunch unsuitable for collider applications. To match the slope and maximum value of the trapezoidal profile, the plasma wave in the second stage must have $E_t/\chi=0.58$ [determined by the slope of $\Lambda$ in Fig. \ref{fig:FF vs conv}(h)], leading to $R_b = 1.4$ as prescribed by Eq. \eqref{eq: beamload correction}. Given $R_b$ and an electron beam size $\sigma_z$, the charge of the drive beam $e N_b$ can be found from the matched condition $R_b=2\sigma_r(n_b/n_0)^{1/2}= (2/\pi^3)^{1/4}[N_b e^2/(c^2 m_e \varepsilon_0  \sigma_z)]^{1/2}$ \cite{tzoufras2009POP}. Using the loaded field of $ 10 \;\mathrm{GV/m}$ and $E_t/\chi=0.58$, the background electron density can be written as $n_0 =10^9\varepsilon_0 /(m_e c^2 E_t^2) =3.2 \times10^{16}/\chi^2\;\mathrm{cm^{-3}}$. The density is selected to ensure that the length of the accelerating structure exceeds the length of the injected bunch. Here, the accelerating structure is set to three times the bunch length, resulting in a background electron density of $n_0 = 5 \times 10^{16} \; \mathrm{cm}^{-3}$. The drive beam parameters utilized in this stage are based on the FACET facility\cite{hogan2010plasma,frederico2011facet}. These and the remaining parameters for the simulations of the acceleration stage are provided in Table \ref{tab:acc_stage_Parameters}.

As shown in Figs. \ref{fig:beam energy}(a,b), the electron beam drives a nonlinear plasma wave with a longitudinal field that is flattened by the trailing bunch. The trapezoidal profile of the bunch and flattened field are preserved over the entire 2 m length of the accelerator [Fig. \ref{fig:beam energy}(c)]. The electron bunch reaches an average energy of 24 GeV with a spread of well less than 1\% [Fig. \ref{fig:beam energy}(d)]. The final slice emittances are nearly uniform across the bunch, and the average emittances, $\varepsilon_x=189$ nm rad and $\varepsilon_y=80$ nm rad, are only slightly larger than their initial values. The accelerator efficiency, i.e., the ratio of energy gained by the bunch to the energy lost by the drive beam, is 43\%. The 24 GeV average energy of the bunch compares favorably to the 28 GeV maximum that could be obtained from the drive beam. These results demonstrate that the flying focus photoinjector can provide the bunch needed for an intermediate energy collider as defined by the Snowmass parameter set \cite{benedetti2022whitepaper}.

\section{CONCLUSIONs AND OUTLOOK}
The laser system suggested for the plasma photoinjector introduced in this paper is consistent with the near-term upgrades planned for the Accelerator Test Facility (ATF) at Brookhaven National Laboratory. Currently, ATF is the only facility worldwide operating a long-wave infrared laser system capable of generating multi-terawatt peak power in clean picosecond pulses. The laser currently produces 2 ps pulses with up to 5 TW of peak power. An active R\&D effort is underway to increase the power to $>20$ TW while reducing the pulse duration to $<500$ fs. This will be achieved by upgrading the laser seed in the front end and through novel methods for temporal compression, such as self-phase modulation in a nonlinear medium. This particular method was recently used to generate a CO$_2$ pulse as short as 675 fs \cite{Pogorelsky2024}, demonstrating a key technology for achieving sub-ps CO$_2$ pulses.

A novel plasma photoinjector that leverages the dynamic focal point of a flying focus pulse can produce electron bunches with the charge, emittance, and luminosity required for collider applications. The moving ionization front driven by the flying focus allows for the injection and formation of an electron bunch with a trapezoidal profile that is nearly ideal for flattening the longitudinal field of a nonlinear plasma wave. In simulation that demonstrated the creation of such a bunch and its subsequent acceleration in an electron beam-driven plasma wave, the bunch achieved a final energy spread of less than 1\% while extracting energy from the driver at $>40 \%$ efficiency. The simulated parameters were targeted towards a mid-scale 20 GeV electron beam based on the requirements of recent Snowmass studies. By adjusting the spot size of the flying focus, the injected trapezoidal bunch can be tailored to have $>1$ nC of charge or $<10$ nm of emittance. This is the first application of spatiotemporal pulse shaping to the creation of structured electron bunches for plasma-based acceleration. Beyond collider applications, the high-quality bunches produced by the flying focus photoinjector also satisfy many of the requirements for advanced radiation sources, such as free-electron lasers. 

Finally, while the parameters presented here utilize the unique capabilities of a long-wavelength driver, the underlying physics of the flying focus injection scheme is broadly applicable. The CO$_2$ laser platform offers distinct advantages, specifically the ability to generate large plasma bubbles and access the blowout regime at lower plasma densities, which relaxes experimental tolerances such as timing jitter. However, with the increasing availability of Petawatt-class Near-Infrared (NIR) laser facilities, similar "macroscopic" wakefield regimes are becoming accessible at shorter wavelengths. Consequently, this injection method is not limited to a specific facility; it can be optimally designed for NIR drivers and implemented across various configurations, including single-stage LWFA, PWFA, or multi-stage architectures, to achieve collider-quality beams.

\section*{Acknowledgments}
We acknowledge the support by the U.S. Department of Energy, Office of Science under Award No. DE-SC-0014043, DE-SC-0020396, DE-SC-0024277, NSF CAREER Award PHY-2238840, resources of NERSC facility, operated under contract No. DE-AC02-5CH11231, and SEAWULF at Stony Brook University. The work of J.P.P. and T.T.S. was supported by the Office of Fusion Energy Sciences under Award Number DE-SC0021057.
\setcounter{table}{0}
\renewcommand{\thetable}{A\arabic{table}}
\renewcommand{\theequation}{A\arabic{equation}}
\renewcommand\thesubsection{\arabic{subsection}}

\section*{APPENDIX A: METHODOLOGY}\label{Methods}

\subsection{Injection stage simulations}
The simulations of the injection stage were performed using the particle-in-cell (PIC) code OSIRIS \cite{Fonseca2002}, which solves the fully relativistic equations of motion for particles and employs a finite-difference time-domain solver for the electromagnetic fields.  The simulations were discretized in the quasi-3D geometry, which represents quantities as a truncated expansion in azimuthal modes \cite{Davidson2015}. For systems with a high degree of cylindrical symmetry, this significantly reduces the computational resources required, while retaining important 3D effects. The customized solver introduced in Ref. \cite{LI2021107580} was used to ensure accurate dispersion of the electromagnetic waves (e.g., the injector and drive pulses). 
The simulations used a moving window propagating at the vacuum speed of light with a size of 400 $\mu$m $\times$ 225 $\mu$m ($20000 \times 2000$ cells) in $z$ and $r$, respectively. Two azimuthal modes ($m = 0, 1$) were sufficient to model the laser pulse and plasma geometry. The initial $(z, r, \theta)$ particle arrangement in each cell was (2,2,8) for a total of 32 particles per cell.

\begin{table}[t!]
\caption{Threshold amplitude ($a_0$) for fully ionizing a particular state of Krypton according to ADK the tunneling ionization model \cite{bruhwiler2003particle}.}
\label{tab:Ionization Table}
 \begin{tabular}{c c c}
 \hline
 \hline
 Ionization levels of Kr & $a_0$($\lambda= 9.2$ $\mu$m ) & $a_0$($\lambda= 0.4$ $\mu$m )\\ 
 Kr$^{7+} \rightarrow$ Kr$^{8+}$ & 1.87& 0.06 \\
 Kr$^{8+} \rightarrow$ Kr$^{9+}$ & 3.10 & 0.16\\
 Kr$^{9+} \rightarrow$ Kr$^{10+}$ & 3.78 & 0.20 \\
 Kr$^{10+} \rightarrow$ Kr$^{11+}$ & 4.51 & 0.23 \\[1ex]
 \hline
 \hline
\end{tabular}
\end{table}

The initial plasma was preionized Kr$^{8+}$, and the injector pulses further ionized the plasma from Kr$^{8+}$ to Kr$^{9+}$. The ionization rates were calculated using the ADK ionization model \cite{bruhwiler2003particle}. The treshold values of $a_0$ for some of the relevant ionization states are displayed in Table \ref{tab:Ionization Table}. The $a_0$ of the CO$_2$ laser pulse was chosen to be between the thresholds of the $8^{th}$ and the $9^{th}$ ionization states.

\subsection{Acceleration stage simulations}
For the acceleration stage, the injected electron bunches resulting from OSIRIS were imported into QPAD. QPAD is a quasi-static PIC code that employs azimuthal decomposition \cite{LI2021107784}. QPAD was used for the acceleration stage because running OSIRIS for a 2 m long plasma would be computationally expensive and yield nearly identical results. The simulations used a moving window propagating at the vacuum speed of light with a size of 261 $\mu$m $\times$ 95 $\mu$m ($2000 \times 1200$ cells) in $\xi$ and $r$, respectively. A single azimuthal mode ($m = 0$) was sufficient for the electron beam, injected bunch, and plasma configuration. The equations of motion for the relativistic electron beam and bunch were evolved with a time step of $dt= 160$ fs.

Unless units are explicitly written or otherwise stated, time and length are normalized to $\omega_p^{-1}$ and $k_p^{-1} =c/\omega_p$, respectively, velocity to the vacuum speed of light $c$, density to the electron density $n_0$, charge to the fundamental charge $e$, potentials to $m_ec^2/e$, linear charge density to $m_e c^2 \varepsilon_0/e$,  and fields to $m_ec\omega_p /e$. Here, $\omega_p=(n_0 e^2/m_e \varepsilon_0)^{1/2}$ is the plasma frequency and $m_e$ is the electron mass. Note that $n_0$ is different in the injection and acceleration stages.

\subsection{Modeling flying focus laser pulses in PIC simulations}
The framework of arbitrarily structured laser (ASTRL) pulses \cite{pierce2023arbitrarily} was used to initialize the flying focus pulse in the injection stage simulations. The transverse electric field of the pulse was initialized as a carrier wave modulating an envelope: $E_\perp ( \mathbf x,t) = \tfrac{1}{2}A(\mathbf x,t) \exp( i k_0 \xi ) \hat {\boldsymbol \varepsilon} + \mathrm{c.c.} $, where $k_0 = 2\pi/\lambda$ is the carrier wavenumber, $\hat {\boldsymbol \varepsilon}$ is the polarization vector, and $A(\mathbf x,t)$ is the envelope. A vacuum paraxial solution was used for the envelope:
\begin{equation}
    A( \mathbf x, t) =  \left(\frac{k_0}{k_p}\right)a_0(\xi) e^{i \beta \xi^2} \left[ \frac{Z_R}{q_z(\xi)} \right] \exp \left[ - \frac{ik(\xi) x_\perp^2}{2q_z(\xi)} \right],
    \label{eq:astrl-analytic}
\end{equation}
where $a_0(\xi)$ is the normalized vector potential, $\beta$ quantifies the amount of chirp, $k(\xi) = k_0 + 2\beta \xi$ is the local frequency, $Z_R = \frac{1}{2} k_0 w_0^2 $ is Rayleigh range, $w_0$ is the focused spot size, $q_z(\xi) = z - f_0(\xi) + i Z_R $ is the complex beam parameter, and $f_0(\xi)$ is the $\xi$-dependent focal length. Equation \ref{eq:astrl-analytic} captures the essential features of a chromatic flying focus \cite{froula2018spatiotemporal,palastro2018ionization}, including the extended focal region, controllable velocity, and chirp. The intensity peak of the pulse can be programmed to move at a constant velocity $v_F$ in vacuum by choosing  $f_0(\xi) = \left[ v_F / ( v_F-1 ) \right] \xi $.
This produces an extended focal region of length $L = | v_F / ( v_F-1 )| \tau$, where $\tau$ is the duration of the pulse set by $a_0(\xi)$. The full-width-at-half-maximum duration of the moving intensity peak within the focal region is $\tau_F = 2 |1-1/v_F| Z_R$. Note that Eqn. \ref{eq:astrl-analytic} is an ASTRL pulse because it is the result of evaluating Eqn. 7 in Ref. \cite{pierce2023arbitrarily} when $B_\eta(\xi)$ is a Dirac delta function and $C_\eta(\mathbf x_\perp, s )$ is a Gaussian beam. 
A special case of this form was used to model a flying focus in Ref. \cite{franke2021optical}.

\setcounter{equation}{0}
\renewcommand{\theequation}{B\arabic{equation}} 
\section*{Appendix B: Transverse emittances} 
According to Refs. \cite{schroeder2015ultra,Xu2014PRL,xu2020phase}, ionization injection with a laser pulse polarized in the $x$-direction results in the saturated transverse emittances
\begin{equation}
\begin{split}
\label{eqn: emittance}
\hspace{-64pt} &\varepsilon_{x} = \frac{\sigma_{x_0}^2}{\sqrt{6}} + \frac{\sqrt{2}\sigma_{px_0}^2 }{\sqrt{3}} \\
\hspace{-64pt} &\varepsilon_{y} = \frac{\sigma_{x_0}^2}{\sqrt{6}},
\end{split}
\end{equation}
where
\begin{equation}
\begin{split}
&\sigma_{x_0} = \frac{w_0} {2\sqrt{2}}, \\
&\sigma_{px_0} = \left(\frac{3\pi r_e}{\alpha^4 \lambda}\right)^{1/2} \left(\frac{U_H}{U_i}\right)^{3/4} \sqrt{2} a_T^{3/2},
\end{split}
\end{equation}
$a_T$ is the threshold amplitude for fully ionizing Kr$^{8+} \rightarrow$ Kr$^{9+}$, $U_i=230$ eV is the ionization potential for Kr$^{8+} \rightarrow$ Kr$^{9+}$, $U_H = 13.6$ eV, $r_e$ is the classical electron radius, and $\alpha$ is the fine structure constant.


\begin{thebibliography}{45}%
\makeatletter
\providecommand \@ifxundefined [1]{%
 \@ifx{#1\undefined}
}%
\providecommand \@ifnum [1]{%
 \ifnum #1\expandafter \@firstoftwo
 \else \expandafter \@secondoftwo
 \fi
}%
\providecommand \@ifx [1]{%
 \ifx #1\expandafter \@firstoftwo
 \else \expandafter \@secondoftwo
 \fi
}%
\providecommand \natexlab [1]{#1}%
\providecommand \enquote  [1]{``#1''}%
\providecommand \bibnamefont  [1]{#1}%
\providecommand \bibfnamefont [1]{#1}%
\providecommand \citenamefont [1]{#1}%
\providecommand \href@noop [0]{\@secondoftwo}%
\providecommand \href [0]{\begingroup \@sanitize@url \@href}%
\providecommand \@href[1]{\@@startlink{#1}\@@href}%
\providecommand \@@href[1]{\endgroup#1\@@endlink}%
\providecommand \@sanitize@url [0]{\catcode `\\12\catcode `\$12\catcode `\&12\catcode `\#12\catcode `\^12\catcode `\_12\catcode `\%12\relax}%
\providecommand \@@startlink[1]{}%
\providecommand \@@endlink[0]{}%
\providecommand \url  [0]{\begingroup\@sanitize@url \@url }%
\providecommand \@url [1]{\endgroup\@href {#1}{\urlprefix }}%
\providecommand \urlprefix  [0]{URL }%
\providecommand \Eprint [0]{\href }%
\providecommand \doibase [0]{https://doi.org/}%
\providecommand \selectlanguage [0]{\@gobble}%
\providecommand \bibinfo  [0]{\@secondoftwo}%
\providecommand \bibfield  [0]{\@secondoftwo}%
\providecommand \translation [1]{[#1]}%
\providecommand \BibitemOpen [0]{}%
\providecommand \bibitemStop [0]{}%
\providecommand \bibitemNoStop [0]{.\EOS\space}%
\providecommand \EOS [0]{\spacefactor3000\relax}%
\providecommand \BibitemShut  [1]{\csname bibitem#1\endcsname}%
\let\auto@bib@innerbib\@empty
\bibitem [{\citenamefont {Mangles}\ \emph {et~al.}(2004)\citenamefont {Mangles}, \citenamefont {Murphy}, \citenamefont {Najmudin}, \citenamefont {Thomas}, \citenamefont {Collier}, \citenamefont {Dangor}, \citenamefont {Divall}, \citenamefont {Foster}, \citenamefont {Gallacher}, \citenamefont {Hooker}, \citenamefont {Jaroszynski}, \citenamefont {Langley}, \citenamefont {Mori}, \citenamefont {Norreys}, \citenamefont {Tsung}, \citenamefont {Viskup}, \citenamefont {Walton},\ and\ \citenamefont {Krushelnick}}]{Mangles04}%
  \BibitemOpen
  \bibfield  {author} {\bibinfo {author} {\bibfnamefont {S.~P.~D.}\ \bibnamefont {Mangles}}, \bibinfo {author} {\bibfnamefont {C.~D.}\ \bibnamefont {Murphy}}, \bibinfo {author} {\bibfnamefont {Z.}~\bibnamefont {Najmudin}}, \bibinfo {author} {\bibfnamefont {A.~G.~R.}\ \bibnamefont {Thomas}}, \bibinfo {author} {\bibfnamefont {J.~L.}\ \bibnamefont {Collier}}, \bibinfo {author} {\bibfnamefont {A.~E.}\ \bibnamefont {Dangor}}, \bibinfo {author} {\bibfnamefont {E.~J.}\ \bibnamefont {Divall}}, \bibinfo {author} {\bibfnamefont {P.~S.}\ \bibnamefont {Foster}}, \bibinfo {author} {\bibfnamefont {J.~G.}\ \bibnamefont {Gallacher}}, \bibinfo {author} {\bibfnamefont {C.~J.}\ \bibnamefont {Hooker}}, \bibinfo {author} {\bibfnamefont {D.~A.}\ \bibnamefont {Jaroszynski}}, \bibinfo {author} {\bibfnamefont {A.~J.}\ \bibnamefont {Langley}}, \bibinfo {author} {\bibfnamefont {W.~B.}\ \bibnamefont {Mori}}, \bibinfo {author} {\bibfnamefont {P.~A.}\ \bibnamefont {Norreys}}, \bibinfo {author} {\bibfnamefont {F.~S.}\ \bibnamefont
  {Tsung}}, \bibinfo {author} {\bibfnamefont {R.}~\bibnamefont {Viskup}}, \bibinfo {author} {\bibfnamefont {B.~R.}\ \bibnamefont {Walton}},\ and\ \bibinfo {author} {\bibfnamefont {K.}~\bibnamefont {Krushelnick}},\ }\bibfield  {title} {\bibinfo {title} {Monoenergetic beams of relativistic electrons from intense laser–plasma interactions},\ }\href {https://doi.org/10.1038/nature02939} {\bibfield  {journal} {\bibinfo  {journal} {Nature}\ }\textbf {\bibinfo {volume} {431}},\ \bibinfo {pages} {535} (\bibinfo {year} {2004})}\BibitemShut {NoStop}%
\bibitem [{\citenamefont {Geddes}\ \emph {et~al.}(2004)\citenamefont {Geddes}, \citenamefont {Toth}, \citenamefont {van Tilborg}, \citenamefont {Esarey}, \citenamefont {Schroeder}, \citenamefont {Bruhwiler}, \citenamefont {Nieter}, \citenamefont {Cary},\ and\ \citenamefont {Leemans}}]{Geddes04}%
  \BibitemOpen
  \bibfield  {author} {\bibinfo {author} {\bibfnamefont {C.~G.~R.}\ \bibnamefont {Geddes}}, \bibinfo {author} {\bibfnamefont {C.}~\bibnamefont {Toth}}, \bibinfo {author} {\bibfnamefont {J.}~\bibnamefont {van Tilborg}}, \bibinfo {author} {\bibfnamefont {E.}~\bibnamefont {Esarey}}, \bibinfo {author} {\bibfnamefont {C.~B.}\ \bibnamefont {Schroeder}}, \bibinfo {author} {\bibfnamefont {D.}~\bibnamefont {Bruhwiler}}, \bibinfo {author} {\bibfnamefont {C.}~\bibnamefont {Nieter}}, \bibinfo {author} {\bibfnamefont {J.}~\bibnamefont {Cary}},\ and\ \bibinfo {author} {\bibfnamefont {W.~P.}\ \bibnamefont {Leemans}},\ }\bibfield  {title} {\bibinfo {title} {High-quality electron beams from a laser wakefield accelerator using plasma-channel guiding},\ }\href {https://doi.org/10.1038/nature02900} {\bibfield  {journal} {\bibinfo  {journal} {Nature}\ }\textbf {\bibinfo {volume} {431}},\ \bibinfo {pages} {538} (\bibinfo {year} {2004})}\BibitemShut {NoStop}%
\bibitem [{\citenamefont {Faure}\ \emph {et~al.}(2004)\citenamefont {Faure}, \citenamefont {Glinec}, \citenamefont {Pukhov}, \citenamefont {Kiselev}, \citenamefont {Gordienko}, \citenamefont {Lefebvre}, \citenamefont {Rousseau}, \citenamefont {Burgy},\ and\ \citenamefont {Malka}}]{Faure04}%
  \BibitemOpen
  \bibfield  {author} {\bibinfo {author} {\bibfnamefont {J.}~\bibnamefont {Faure}}, \bibinfo {author} {\bibfnamefont {Y.}~\bibnamefont {Glinec}}, \bibinfo {author} {\bibfnamefont {A.}~\bibnamefont {Pukhov}}, \bibinfo {author} {\bibfnamefont {S.}~\bibnamefont {Kiselev}}, \bibinfo {author} {\bibfnamefont {S.}~\bibnamefont {Gordienko}}, \bibinfo {author} {\bibfnamefont {E.}~\bibnamefont {Lefebvre}}, \bibinfo {author} {\bibfnamefont {J.-P.}\ \bibnamefont {Rousseau}}, \bibinfo {author} {\bibfnamefont {F.}~\bibnamefont {Burgy}},\ and\ \bibinfo {author} {\bibfnamefont {V.}~\bibnamefont {Malka}},\ }\bibfield  {title} {\bibinfo {title} {A laser–plasma accelerator producing monoenergetic electron beams},\ }\href {https://doi.org/10.1038/nature02963} {\bibfield  {journal} {\bibinfo  {journal} {Nature}\ }\textbf {\bibinfo {volume} {431}},\ \bibinfo {pages} {541} (\bibinfo {year} {2004})}\BibitemShut {NoStop}%
\bibitem [{\citenamefont {Schroeder}\ \emph {et~al.}(2023)\citenamefont {Schroeder}, \citenamefont {Albert}, \citenamefont {Benedetti}, \citenamefont {Bromage}, \citenamefont {Bruhwiler}, \citenamefont {Bulanov}, \citenamefont {Campbell}, \citenamefont {Cook}, \citenamefont {Cros}, \citenamefont {Downer}, \citenamefont {Esarey}, \citenamefont {Froula}, \citenamefont {Fuchs}, \citenamefont {Geddes}, \citenamefont {Gessner}, \citenamefont {Gonsalves}, \citenamefont {Hogan}, \citenamefont {Hooker}, \citenamefont {Huebl}, \citenamefont {Jing}, \citenamefont {Joshi}, \citenamefont {Krushelnick}, \citenamefont {Leemans}, \citenamefont {Lehe}, \citenamefont {Maier}, \citenamefont {Milchberg}, \citenamefont {Mori}, \citenamefont {Nakamura}, \citenamefont {Osterhoff}, \citenamefont {Palastro}, \citenamefont {Palmer}, \citenamefont {Põder}, \citenamefont {Power}, \citenamefont {Shadwick}, \citenamefont {Terzani}, \citenamefont {Thévenet}, \citenamefont {Thomas}, \citenamefont {van Tilborg}, \citenamefont {Turner},
  \citenamefont {Vafaei-Najafabadi}, \citenamefont {Vay}, \citenamefont {Zhou},\ and\ \citenamefont {Zuegel}}]{benedetti2022linear}%
  \BibitemOpen
  \bibfield  {author} {\bibinfo {author} {\bibfnamefont {C.}~\bibnamefont {Schroeder}}, \bibinfo {author} {\bibfnamefont {F.}~\bibnamefont {Albert}}, \bibinfo {author} {\bibfnamefont {C.}~\bibnamefont {Benedetti}}, \bibinfo {author} {\bibfnamefont {J.}~\bibnamefont {Bromage}}, \bibinfo {author} {\bibfnamefont {D.}~\bibnamefont {Bruhwiler}}, \bibinfo {author} {\bibfnamefont {S.}~\bibnamefont {Bulanov}}, \bibinfo {author} {\bibfnamefont {E.}~\bibnamefont {Campbell}}, \bibinfo {author} {\bibfnamefont {N.}~\bibnamefont {Cook}}, \bibinfo {author} {\bibfnamefont {B.}~\bibnamefont {Cros}}, \bibinfo {author} {\bibfnamefont {M.}~\bibnamefont {Downer}}, \bibinfo {author} {\bibfnamefont {E.}~\bibnamefont {Esarey}}, \bibinfo {author} {\bibfnamefont {D.}~\bibnamefont {Froula}}, \bibinfo {author} {\bibfnamefont {M.}~\bibnamefont {Fuchs}}, \bibinfo {author} {\bibfnamefont {C.}~\bibnamefont {Geddes}}, \bibinfo {author} {\bibfnamefont {S.}~\bibnamefont {Gessner}}, \bibinfo {author} {\bibfnamefont {A.}~\bibnamefont
  {Gonsalves}}, \bibinfo {author} {\bibfnamefont {M.}~\bibnamefont {Hogan}}, \bibinfo {author} {\bibfnamefont {S.}~\bibnamefont {Hooker}}, \bibinfo {author} {\bibfnamefont {A.}~\bibnamefont {Huebl}}, \bibinfo {author} {\bibfnamefont {C.}~\bibnamefont {Jing}}, \bibinfo {author} {\bibfnamefont {C.}~\bibnamefont {Joshi}}, \bibinfo {author} {\bibfnamefont {K.}~\bibnamefont {Krushelnick}}, \bibinfo {author} {\bibfnamefont {W.}~\bibnamefont {Leemans}}, \bibinfo {author} {\bibfnamefont {R.}~\bibnamefont {Lehe}}, \bibinfo {author} {\bibfnamefont {A.}~\bibnamefont {Maier}}, \bibinfo {author} {\bibfnamefont {H.}~\bibnamefont {Milchberg}}, \bibinfo {author} {\bibfnamefont {W.}~\bibnamefont {Mori}}, \bibinfo {author} {\bibfnamefont {K.}~\bibnamefont {Nakamura}}, \bibinfo {author} {\bibfnamefont {J.}~\bibnamefont {Osterhoff}}, \bibinfo {author} {\bibfnamefont {J.}~\bibnamefont {Palastro}}, \bibinfo {author} {\bibfnamefont {M.}~\bibnamefont {Palmer}}, \bibinfo {author} {\bibfnamefont {K.}~\bibnamefont {Põder}}, \bibinfo
  {author} {\bibfnamefont {J.}~\bibnamefont {Power}}, \bibinfo {author} {\bibfnamefont {B.}~\bibnamefont {Shadwick}}, \bibinfo {author} {\bibfnamefont {D.}~\bibnamefont {Terzani}}, \bibinfo {author} {\bibfnamefont {M.}~\bibnamefont {Thévenet}}, \bibinfo {author} {\bibfnamefont {A.}~\bibnamefont {Thomas}}, \bibinfo {author} {\bibfnamefont {J.}~\bibnamefont {van Tilborg}}, \bibinfo {author} {\bibfnamefont {M.}~\bibnamefont {Turner}}, \bibinfo {author} {\bibfnamefont {N.}~\bibnamefont {Vafaei-Najafabadi}}, \bibinfo {author} {\bibfnamefont {J.-L.}\ \bibnamefont {Vay}}, \bibinfo {author} {\bibfnamefont {T.}~\bibnamefont {Zhou}},\ and\ \bibinfo {author} {\bibfnamefont {J.}~\bibnamefont {Zuegel}},\ }\bibfield  {title} {\bibinfo {title} {Linear colliders based on laser-plasma accelerators},\ }\href {https://doi.org/10.1088/1748-0221/18/06/T06001} {\bibfield  {journal} {\bibinfo  {journal} {Journal of Instrumentation}\ }\textbf {\bibinfo {volume} {18}}\bibinfo  {number} { (06)},\ \bibinfo {pages}
  {T06001}}\BibitemShut {NoStop}%
\bibitem [{\citenamefont {Bulanov}\ \emph {et~al.}(2024)\citenamefont {Bulanov}, \citenamefont {Aidala}, \citenamefont {Benedetti}, \citenamefont {Bernstein}, \citenamefont {Esarey}, \citenamefont {Geddes}, \citenamefont {Gessner}, \citenamefont {Gonsalves}, \citenamefont {Hogan}, \citenamefont {Jacobs} \emph {et~al.}}]{benedetti2022whitepaper}%
  \BibitemOpen
\bibfield  {number} {  }\bibfield  {author} {\bibinfo {author} {\bibfnamefont {S.}~\bibnamefont {Bulanov}}, \bibinfo {author} {\bibfnamefont {C.}~\bibnamefont {Aidala}}, \bibinfo {author} {\bibfnamefont {C.}~\bibnamefont {Benedetti}}, \bibinfo {author} {\bibfnamefont {R.}~\bibnamefont {Bernstein}}, \bibinfo {author} {\bibfnamefont {E.}~\bibnamefont {Esarey}}, \bibinfo {author} {\bibfnamefont {C.}~\bibnamefont {Geddes}}, \bibinfo {author} {\bibfnamefont {S.}~\bibnamefont {Gessner}}, \bibinfo {author} {\bibfnamefont {A.}~\bibnamefont {Gonsalves}}, \bibinfo {author} {\bibfnamefont {M.}~\bibnamefont {Hogan}}, \bibinfo {author} {\bibfnamefont {P.}~\bibnamefont {Jacobs}}, \emph {et~al.},\ }\bibfield  {title} {\bibinfo {title} {The science case for an intermediate energy advanced and novel accelerator linear collider facility},\ }\href {https://doi.org/10.1088/1748-0221/19/01/T01010} {\bibfield  {journal} {\bibinfo  {journal} {Journal of Instrumentation}\ }\textbf {\bibinfo {volume} {19}}\bibinfo  {number} { (01)},\
  \bibinfo {pages} {T01010}}\BibitemShut {NoStop}%
\bibitem [{\citenamefont {Faure}(2017)}]{faure2017plasma}%
  \BibitemOpen
\bibfield  {number} {  }\bibfield  {author} {\bibinfo {author} {\bibfnamefont {J.}~\bibnamefont {Faure}},\ }\href {https://doi.org/10.5170/CERN-2016-001.143} {\emph {\bibinfo {title} {Plasma Injection Schemes for Laser-Plasma Accelerators}}},\ \bibinfo {type} {Tech. Rep.}\ (\bibinfo  {institution} {CERN},\ \bibinfo {year} {2017})\BibitemShut {NoStop}%
\bibitem [{\citenamefont {Fuchs}\ \emph {et~al.}(2024)\citenamefont {Fuchs}, \citenamefont {Andonian}, \citenamefont {Apsimon}, \citenamefont {B{\"u}scher}, \citenamefont {Downer}, \citenamefont {Filippetto}, \citenamefont {Lehrach}, \citenamefont {Schroeder}, \citenamefont {Shadwick}, \citenamefont {Thomas} \emph {et~al.}}]{fuchs2022snowmass}%
  \BibitemOpen
  \bibfield  {author} {\bibinfo {author} {\bibfnamefont {M.}~\bibnamefont {Fuchs}}, \bibinfo {author} {\bibfnamefont {G.}~\bibnamefont {Andonian}}, \bibinfo {author} {\bibfnamefont {O.}~\bibnamefont {Apsimon}}, \bibinfo {author} {\bibfnamefont {M.}~\bibnamefont {B{\"u}scher}}, \bibinfo {author} {\bibfnamefont {M.}~\bibnamefont {Downer}}, \bibinfo {author} {\bibfnamefont {D.}~\bibnamefont {Filippetto}}, \bibinfo {author} {\bibfnamefont {A.}~\bibnamefont {Lehrach}}, \bibinfo {author} {\bibfnamefont {C.}~\bibnamefont {Schroeder}}, \bibinfo {author} {\bibfnamefont {B.}~\bibnamefont {Shadwick}}, \bibinfo {author} {\bibfnamefont {A.}~\bibnamefont {Thomas}}, \emph {et~al.},\ }\bibfield  {title} {\bibinfo {title} {Plasma-based particle sources},\ }\href {https://doi.org/10.1088/1748-0221/19/01/T01004} {\bibfield  {journal} {\bibinfo  {journal} {Journal of Instrumentation}\ }\textbf {\bibinfo {volume} {19}}\bibinfo  {number} { (01)},\ \bibinfo {pages} {T01004}}\BibitemShut {NoStop}%
\bibitem [{\citenamefont {Froula}\ \emph {et~al.}(2019)\citenamefont {Froula}, \citenamefont {Palastro}, \citenamefont {Turnbull}, \citenamefont {Davies}, \citenamefont {Nguyen}, \citenamefont {Howard}, \citenamefont {Ramsey}, \citenamefont {Franke}, \citenamefont {Bahk}, \citenamefont {Begishev} \emph {et~al.}}]{froula2019flying}%
  \BibitemOpen
\bibfield  {number} {  }\bibfield  {author} {\bibinfo {author} {\bibfnamefont {D.}~\bibnamefont {Froula}}, \bibinfo {author} {\bibfnamefont {J.}~\bibnamefont {Palastro}}, \bibinfo {author} {\bibfnamefont {D.}~\bibnamefont {Turnbull}}, \bibinfo {author} {\bibfnamefont {A.}~\bibnamefont {Davies}}, \bibinfo {author} {\bibfnamefont {L.}~\bibnamefont {Nguyen}}, \bibinfo {author} {\bibfnamefont {A.}~\bibnamefont {Howard}}, \bibinfo {author} {\bibfnamefont {D.}~\bibnamefont {Ramsey}}, \bibinfo {author} {\bibfnamefont {P.}~\bibnamefont {Franke}}, \bibinfo {author} {\bibfnamefont {S.-W.}\ \bibnamefont {Bahk}}, \bibinfo {author} {\bibfnamefont {I.}~\bibnamefont {Begishev}}, \emph {et~al.},\ }\bibfield  {title} {\bibinfo {title} {Flying focus: Spatial and temporal control of intensity for laser-based applications},\ }\href {https://doi.org/10.1063/1.5086308} {\bibfield  {journal} {\bibinfo  {journal} {Physics of Plasmas}\ }\textbf {\bibinfo {volume} {26}},\ \bibinfo {pages} {033110} (\bibinfo {year} {2019})}\BibitemShut
  {NoStop}%
\bibitem [{\citenamefont {Xu}\ \emph {et~al.}(2014{\natexlab{a}})\citenamefont {Xu}, \citenamefont {Wu}, \citenamefont {Zhang}, \citenamefont {Li}, \citenamefont {Wan}, \citenamefont {Hua}, \citenamefont {Pai}, \citenamefont {Lu}, \citenamefont {Yu}, \citenamefont {Joshi} \emph {et~al.}}]{xu2014low}%
  \BibitemOpen
  \bibfield  {author} {\bibinfo {author} {\bibfnamefont {X.}~\bibnamefont {Xu}}, \bibinfo {author} {\bibfnamefont {Y.}~\bibnamefont {Wu}}, \bibinfo {author} {\bibfnamefont {C.}~\bibnamefont {Zhang}}, \bibinfo {author} {\bibfnamefont {F.}~\bibnamefont {Li}}, \bibinfo {author} {\bibfnamefont {Y.}~\bibnamefont {Wan}}, \bibinfo {author} {\bibfnamefont {J.}~\bibnamefont {Hua}}, \bibinfo {author} {\bibfnamefont {C.-H.}\ \bibnamefont {Pai}}, \bibinfo {author} {\bibfnamefont {W.}~\bibnamefont {Lu}}, \bibinfo {author} {\bibfnamefont {P.}~\bibnamefont {Yu}}, \bibinfo {author} {\bibfnamefont {C.}~\bibnamefont {Joshi}}, \emph {et~al.},\ }\bibfield  {title} {\bibinfo {title} {Low emittance electron beam generation from a laser wakefield accelerator using two laser pulses with different wavelengths},\ }\href {https://doi.org/10.1103/PhysRevSTAB.17.061301} {\bibfield  {journal} {\bibinfo  {journal} {Physical Review Special Topics-Accelerators and Beams}\ }\textbf {\bibinfo {volume} {17}},\ \bibinfo {pages} {061301}
  (\bibinfo {year} {2014}{\natexlab{a}})}\BibitemShut {NoStop}%
\bibitem [{\citenamefont {Yu}\ \emph {et~al.}(2014)\citenamefont {Yu}, \citenamefont {Esarey}, \citenamefont {Schroeder}, \citenamefont {Vay}, \citenamefont {Benedetti}, \citenamefont {Geddes}, \citenamefont {Chen},\ and\ \citenamefont {Leemans}}]{Schroeder2014}%
  \BibitemOpen
  \bibfield  {author} {\bibinfo {author} {\bibfnamefont {L.-L.}\ \bibnamefont {Yu}}, \bibinfo {author} {\bibfnamefont {E.}~\bibnamefont {Esarey}}, \bibinfo {author} {\bibfnamefont {C.~B.}\ \bibnamefont {Schroeder}}, \bibinfo {author} {\bibfnamefont {J.-L.}\ \bibnamefont {Vay}}, \bibinfo {author} {\bibfnamefont {C.}~\bibnamefont {Benedetti}}, \bibinfo {author} {\bibfnamefont {C.~G.~R.}\ \bibnamefont {Geddes}}, \bibinfo {author} {\bibfnamefont {M.}~\bibnamefont {Chen}},\ and\ \bibinfo {author} {\bibfnamefont {W.~P.}\ \bibnamefont {Leemans}},\ }\bibfield  {title} {\bibinfo {title} {Two-color laser-ionization injection},\ }\href {https://doi.org/10.1103/PhysRevLett.112.125001} {\bibfield  {journal} {\bibinfo  {journal} {Phys. Rev. Lett.}\ }\textbf {\bibinfo {volume} {112}},\ \bibinfo {pages} {125001} (\bibinfo {year} {2014})}\BibitemShut {NoStop}%
\bibitem [{\citenamefont {Schroeder}\ \emph {et~al.}(2015)\citenamefont {Schroeder}, \citenamefont {Benedetti}, \citenamefont {Bulanov}, \citenamefont {Chen}, \citenamefont {Esarey}, \citenamefont {Geddes}, \citenamefont {Vay}, \citenamefont {Yu},\ and\ \citenamefont {Leemans}}]{schroeder2015ultra}%
  \BibitemOpen
  \bibfield  {author} {\bibinfo {author} {\bibfnamefont {C.}~\bibnamefont {Schroeder}}, \bibinfo {author} {\bibfnamefont {C.}~\bibnamefont {Benedetti}}, \bibinfo {author} {\bibfnamefont {S.}~\bibnamefont {Bulanov}}, \bibinfo {author} {\bibfnamefont {M.}~\bibnamefont {Chen}}, \bibinfo {author} {\bibfnamefont {E.}~\bibnamefont {Esarey}}, \bibinfo {author} {\bibfnamefont {C.~C.}\ \bibnamefont {Geddes}}, \bibinfo {author} {\bibfnamefont {J.-L.}\ \bibnamefont {Vay}}, \bibinfo {author} {\bibfnamefont {L.}~\bibnamefont {Yu}},\ and\ \bibinfo {author} {\bibfnamefont {W.}~\bibnamefont {Leemans}},\ }\bibfield  {title} {\bibinfo {title} {Ultra-low emittance beam generation using two-color ionization injection in laser-plasma accelerators},\ }in\ \href {https://doi.org/10.1117/12.2182035} {\emph {\bibinfo {booktitle} {Laser Acceleration of Electrons, Protons, and Ions III; and Medical Applications of Laser-Generated Beams of Particles III}}},\ Vol.\ \bibinfo {volume} {9514}\ (\bibinfo {organization} {SPIE},\ \bibinfo
  {year} {2015})\ pp.\ \bibinfo {pages} {8--14}\BibitemShut {NoStop}%
\bibitem [{\citenamefont {Palastro}\ \emph {et~al.}(2020)\citenamefont {Palastro}, \citenamefont {Shaw}, \citenamefont {Franke}, \citenamefont {Ramsey}, \citenamefont {Simpson},\ and\ \citenamefont {Froula}}]{Palastro2020dephasing}%
  \BibitemOpen
  \bibfield  {author} {\bibinfo {author} {\bibfnamefont {J.~P.}\ \bibnamefont {Palastro}}, \bibinfo {author} {\bibfnamefont {J.~L.}\ \bibnamefont {Shaw}}, \bibinfo {author} {\bibfnamefont {P.}~\bibnamefont {Franke}}, \bibinfo {author} {\bibfnamefont {D.}~\bibnamefont {Ramsey}}, \bibinfo {author} {\bibfnamefont {T.~T.}\ \bibnamefont {Simpson}},\ and\ \bibinfo {author} {\bibfnamefont {D.~H.}\ \bibnamefont {Froula}},\ }\bibfield  {title} {\bibinfo {title} {Dephasingless laser wakefield acceleration},\ }\href {https://doi.org/10.1103/PhysRevLett.124.134802} {\bibfield  {journal} {\bibinfo  {journal} {Phys. Rev. Lett.}\ }\textbf {\bibinfo {volume} {124}},\ \bibinfo {pages} {134802} (\bibinfo {year} {2020})}\BibitemShut {NoStop}%
\bibitem [{\citenamefont {Tzoufras}\ \emph {et~al.}(2008)\citenamefont {Tzoufras}, \citenamefont {Lu}, \citenamefont {Tsung}, \citenamefont {Huang}, \citenamefont {Mori}, \citenamefont {Katsouleas}, \citenamefont {Vieira}, \citenamefont {Fonseca},\ and\ \citenamefont {Silva}}]{Tzoufras2008PRL}%
  \BibitemOpen
  \bibfield  {author} {\bibinfo {author} {\bibfnamefont {M.}~\bibnamefont {Tzoufras}}, \bibinfo {author} {\bibfnamefont {W.}~\bibnamefont {Lu}}, \bibinfo {author} {\bibfnamefont {F.~S.}\ \bibnamefont {Tsung}}, \bibinfo {author} {\bibfnamefont {C.}~\bibnamefont {Huang}}, \bibinfo {author} {\bibfnamefont {W.~B.}\ \bibnamefont {Mori}}, \bibinfo {author} {\bibfnamefont {T.}~\bibnamefont {Katsouleas}}, \bibinfo {author} {\bibfnamefont {J.}~\bibnamefont {Vieira}}, \bibinfo {author} {\bibfnamefont {R.~A.}\ \bibnamefont {Fonseca}},\ and\ \bibinfo {author} {\bibfnamefont {L.~O.}\ \bibnamefont {Silva}},\ }\bibfield  {title} {\bibinfo {title} {Beam loading in the nonlinear regime of plasma-based acceleration},\ }\href {https://doi.org/10.1103/PhysRevLett.101.145002} {\bibfield  {journal} {\bibinfo  {journal} {Phys. Rev. Lett.}\ }\textbf {\bibinfo {volume} {101}},\ \bibinfo {pages} {145002} (\bibinfo {year} {2008})}\BibitemShut {NoStop}%
\bibitem [{\citenamefont {Li}\ \emph {et~al.}(2019)\citenamefont {Li}, \citenamefont {Li}, \citenamefont {Ain}, \citenamefont {Hur}, \citenamefont {Ting}, \citenamefont {Kulagin}, \citenamefont {Kamperidis},\ and\ \citenamefont {Hafz}}]{li2019laser}%
  \BibitemOpen
  \bibfield  {author} {\bibinfo {author} {\bibfnamefont {S.}~\bibnamefont {Li}}, \bibinfo {author} {\bibfnamefont {G.}~\bibnamefont {Li}}, \bibinfo {author} {\bibfnamefont {Q.}~\bibnamefont {Ain}}, \bibinfo {author} {\bibfnamefont {M.~S.}\ \bibnamefont {Hur}}, \bibinfo {author} {\bibfnamefont {A.~C.}\ \bibnamefont {Ting}}, \bibinfo {author} {\bibfnamefont {V.~V.}\ \bibnamefont {Kulagin}}, \bibinfo {author} {\bibfnamefont {C.}~\bibnamefont {Kamperidis}},\ and\ \bibinfo {author} {\bibfnamefont {N.~A.}\ \bibnamefont {Hafz}},\ }\bibfield  {title} {\bibinfo {title} {A laser-plasma accelerator driven by two-color relativistic femtosecond laser pulses},\ }\href {https://doi.org/10.1126/sciadv.aav7940} {\bibfield  {journal} {\bibinfo  {journal} {Science Advances}\ }\textbf {\bibinfo {volume} {5}},\ \bibinfo {pages} {eaav7940} (\bibinfo {year} {2019})}\BibitemShut {NoStop}%
\bibitem [{\citenamefont {Lu}\ \emph {et~al.}(2006)\citenamefont {Lu}, \citenamefont {Huang}, \citenamefont {Zhou}, \citenamefont {Mori},\ and\ \citenamefont {Katsouleas}}]{lu2006nonlinear}%
  \BibitemOpen
  \bibfield  {author} {\bibinfo {author} {\bibfnamefont {W.}~\bibnamefont {Lu}}, \bibinfo {author} {\bibfnamefont {C.}~\bibnamefont {Huang}}, \bibinfo {author} {\bibfnamefont {M.}~\bibnamefont {Zhou}}, \bibinfo {author} {\bibfnamefont {W.~B.}\ \bibnamefont {Mori}},\ and\ \bibinfo {author} {\bibfnamefont {T.}~\bibnamefont {Katsouleas}},\ }\bibfield  {title} {\bibinfo {title} {Nonlinear theory for relativistic plasma wakefields in the blowout regime},\ }\href {https://doi.org/10.1103/PhysRevLett.96.165002} {\bibfield  {journal} {\bibinfo  {journal} {Physical Review Letters}\ }\textbf {\bibinfo {volume} {96}},\ \bibinfo {pages} {165002} (\bibinfo {year} {2006})}\BibitemShut {NoStop}%
\bibitem [{\citenamefont {Froula}\ \emph {et~al.}(2018)\citenamefont {Froula}, \citenamefont {Turnbull}, \citenamefont {Davies}, \citenamefont {Kessler}, \citenamefont {Haberberger}, \citenamefont {Palastro}, \citenamefont {Bahk}, \citenamefont {Begishev}, \citenamefont {Boni}, \citenamefont {Bucht} \emph {et~al.}}]{froula2018spatiotemporal}%
  \BibitemOpen
  \bibfield  {author} {\bibinfo {author} {\bibfnamefont {D.~H.}\ \bibnamefont {Froula}}, \bibinfo {author} {\bibfnamefont {D.}~\bibnamefont {Turnbull}}, \bibinfo {author} {\bibfnamefont {A.~S.}\ \bibnamefont {Davies}}, \bibinfo {author} {\bibfnamefont {T.~J.}\ \bibnamefont {Kessler}}, \bibinfo {author} {\bibfnamefont {D.}~\bibnamefont {Haberberger}}, \bibinfo {author} {\bibfnamefont {J.~P.}\ \bibnamefont {Palastro}}, \bibinfo {author} {\bibfnamefont {S.-W.}\ \bibnamefont {Bahk}}, \bibinfo {author} {\bibfnamefont {I.~A.}\ \bibnamefont {Begishev}}, \bibinfo {author} {\bibfnamefont {R.}~\bibnamefont {Boni}}, \bibinfo {author} {\bibfnamefont {S.}~\bibnamefont {Bucht}}, \emph {et~al.},\ }\bibfield  {title} {\bibinfo {title} {Spatiotemporal control of laser intensity},\ }\href {https://doi.org/10.1038/s41566-018-0121-8} {\bibfield  {journal} {\bibinfo  {journal} {Nature Photonics}\ }\textbf {\bibinfo {volume} {12}},\ \bibinfo {pages} {262} (\bibinfo {year} {2018})}\BibitemShut {NoStop}%
\bibitem [{\citenamefont {Palastro}\ \emph {et~al.}(2018)\citenamefont {Palastro}, \citenamefont {Turnbull}, \citenamefont {Bahk}, \citenamefont {Follett}, \citenamefont {Shaw}, \citenamefont {Haberberger}, \citenamefont {Bromage},\ and\ \citenamefont {Froula}}]{palastro2018ionization}%
  \BibitemOpen
  \bibfield  {author} {\bibinfo {author} {\bibfnamefont {J.~P.}\ \bibnamefont {Palastro}}, \bibinfo {author} {\bibfnamefont {D.}~\bibnamefont {Turnbull}}, \bibinfo {author} {\bibfnamefont {S.-W.}\ \bibnamefont {Bahk}}, \bibinfo {author} {\bibfnamefont {R.~K.}\ \bibnamefont {Follett}}, \bibinfo {author} {\bibfnamefont {J.~L.}\ \bibnamefont {Shaw}}, \bibinfo {author} {\bibfnamefont {D.}~\bibnamefont {Haberberger}}, \bibinfo {author} {\bibfnamefont {J.}~\bibnamefont {Bromage}},\ and\ \bibinfo {author} {\bibfnamefont {D.~H.}\ \bibnamefont {Froula}},\ }\bibfield  {title} {\bibinfo {title} {Ionization waves of arbitrary velocity driven by a flying focus},\ }\href {https://doi.org/10.1103/PhysRevA.97.033835} {\bibfield  {journal} {\bibinfo  {journal} {Phys. Rev. A}\ }\textbf {\bibinfo {volume} {97}},\ \bibinfo {pages} {033835} (\bibinfo {year} {2018})}\BibitemShut {NoStop}%
\bibitem [{\citenamefont {Turnbull}\ \emph {et~al.}(2018)\citenamefont {Turnbull}, \citenamefont {Franke}, \citenamefont {Katz}, \citenamefont {Palastro}, \citenamefont {Begishev}, \citenamefont {Boni}, \citenamefont {Bromage}, \citenamefont {Milder}, \citenamefont {Shaw},\ and\ \citenamefont {Froula}}]{turnbull2018ionization}%
  \BibitemOpen
  \bibfield  {author} {\bibinfo {author} {\bibfnamefont {D.}~\bibnamefont {Turnbull}}, \bibinfo {author} {\bibfnamefont {P.}~\bibnamefont {Franke}}, \bibinfo {author} {\bibfnamefont {J.}~\bibnamefont {Katz}}, \bibinfo {author} {\bibfnamefont {J.}~\bibnamefont {Palastro}}, \bibinfo {author} {\bibfnamefont {I.}~\bibnamefont {Begishev}}, \bibinfo {author} {\bibfnamefont {R.}~\bibnamefont {Boni}}, \bibinfo {author} {\bibfnamefont {J.}~\bibnamefont {Bromage}}, \bibinfo {author} {\bibfnamefont {A.}~\bibnamefont {Milder}}, \bibinfo {author} {\bibfnamefont {J.}~\bibnamefont {Shaw}},\ and\ \bibinfo {author} {\bibfnamefont {D.}~\bibnamefont {Froula}},\ }\bibfield  {title} {\bibinfo {title} {Ionization waves of arbitrary velocity},\ }\href {https://doi.org/10.1103/PhysRevLett.120.225001} {\bibfield  {journal} {\bibinfo  {journal} {Physical Review Letters}\ }\textbf {\bibinfo {volume} {120}},\ \bibinfo {pages} {225001} (\bibinfo {year} {2018})}\BibitemShut {NoStop}%
\bibitem [{\citenamefont {Palastro}\ \emph {et~al.}(2021)\citenamefont {Palastro}, \citenamefont {Malaca}, \citenamefont {Vieira}, \citenamefont {Ramsey}, \citenamefont {Simpson}, \citenamefont {Franke}, \citenamefont {Shaw},\ and\ \citenamefont {Froula}}]{palastro2021laser}%
  \BibitemOpen
  \bibfield  {author} {\bibinfo {author} {\bibfnamefont {J.}~\bibnamefont {Palastro}}, \bibinfo {author} {\bibfnamefont {B.}~\bibnamefont {Malaca}}, \bibinfo {author} {\bibfnamefont {J.}~\bibnamefont {Vieira}}, \bibinfo {author} {\bibfnamefont {D.}~\bibnamefont {Ramsey}}, \bibinfo {author} {\bibfnamefont {T.}~\bibnamefont {Simpson}}, \bibinfo {author} {\bibfnamefont {P.}~\bibnamefont {Franke}}, \bibinfo {author} {\bibfnamefont {J.}~\bibnamefont {Shaw}},\ and\ \bibinfo {author} {\bibfnamefont {D.}~\bibnamefont {Froula}},\ }\bibfield  {title} {\bibinfo {title} {Laser-plasma acceleration beyond wave breaking},\ }\href {https://doi.org/10.1063/5.0031235} {\bibfield  {journal} {\bibinfo  {journal} {Physics of Plasmas}\ }\textbf {\bibinfo {volume} {28}},\ \bibinfo {pages} {013109} (\bibinfo {year} {2021})}\BibitemShut {NoStop}%
\bibitem [{\citenamefont {Miller}\ \emph {et~al.}(2023)\citenamefont {Miller}, \citenamefont {Pierce}, \citenamefont {Ambat}, \citenamefont {Shaw}, \citenamefont {Weichman}, \citenamefont {Mori}, \citenamefont {Froula},\ and\ \citenamefont {Palastro}}]{miller2023dephasingless}%
  \BibitemOpen
  \bibfield  {author} {\bibinfo {author} {\bibfnamefont {K.~G.}\ \bibnamefont {Miller}}, \bibinfo {author} {\bibfnamefont {J.~R.}\ \bibnamefont {Pierce}}, \bibinfo {author} {\bibfnamefont {M.~V.}\ \bibnamefont {Ambat}}, \bibinfo {author} {\bibfnamefont {J.~L.}\ \bibnamefont {Shaw}}, \bibinfo {author} {\bibfnamefont {K.}~\bibnamefont {Weichman}}, \bibinfo {author} {\bibfnamefont {W.~B.}\ \bibnamefont {Mori}}, \bibinfo {author} {\bibfnamefont {D.~H.}\ \bibnamefont {Froula}},\ and\ \bibinfo {author} {\bibfnamefont {J.~P.}\ \bibnamefont {Palastro}},\ }\bibfield  {title} {\bibinfo {title} {Dephasingless laser wakefield acceleration in the bubble regime},\ }\href {https://doi.org/10.1038/s41598-023-48249-4} {\bibfield  {journal} {\bibinfo  {journal} {Scientific Reports}\ }\textbf {\bibinfo {volume} {13}},\ \bibinfo {pages} {21306} (\bibinfo {year} {2023})}\BibitemShut {NoStop}%
\bibitem [{\citenamefont {Pigeon}\ \emph {et~al.}(2024)\citenamefont {Pigeon}, \citenamefont {Franke}, \citenamefont {Chong}, \citenamefont {Katz}, \citenamefont {Boni}, \citenamefont {Dorrer}, \citenamefont {Palastro},\ and\ \citenamefont {Froula}}]{pigeon2024ultrabroadband}%
  \BibitemOpen
  \bibfield  {author} {\bibinfo {author} {\bibfnamefont {J.}~\bibnamefont {Pigeon}}, \bibinfo {author} {\bibfnamefont {P.}~\bibnamefont {Franke}}, \bibinfo {author} {\bibfnamefont {M.~L.~P.}\ \bibnamefont {Chong}}, \bibinfo {author} {\bibfnamefont {J.}~\bibnamefont {Katz}}, \bibinfo {author} {\bibfnamefont {R.}~\bibnamefont {Boni}}, \bibinfo {author} {\bibfnamefont {C.}~\bibnamefont {Dorrer}}, \bibinfo {author} {\bibfnamefont {J.}~\bibnamefont {Palastro}},\ and\ \bibinfo {author} {\bibfnamefont {D.}~\bibnamefont {Froula}},\ }\bibfield  {title} {\bibinfo {title} {Ultrabroadband flying-focus using an axiparabola-echelon pair},\ }\href {https://doi.org/10.1364/OE.504175} {\bibfield  {journal} {\bibinfo  {journal} {Optics Express}\ }\textbf {\bibinfo {volume} {32}},\ \bibinfo {pages} {576} (\bibinfo {year} {2024})}\BibitemShut {NoStop}%
\bibitem [{\citenamefont {Sainte-Marie}\ \emph {et~al.}(2017)\citenamefont {Sainte-Marie}, \citenamefont {Gobert},\ and\ \citenamefont {Quere}}]{sainte2017controlling}%
  \BibitemOpen
  \bibfield  {author} {\bibinfo {author} {\bibfnamefont {A.}~\bibnamefont {Sainte-Marie}}, \bibinfo {author} {\bibfnamefont {O.}~\bibnamefont {Gobert}},\ and\ \bibinfo {author} {\bibfnamefont {F.}~\bibnamefont {Quere}},\ }\bibfield  {title} {\bibinfo {title} {Controlling the velocity of ultrashort light pulses in vacuum through spatio-temporal couplings},\ }\href {https://doi.org/10.1364/OPTICA.4.001298} {\bibfield  {journal} {\bibinfo  {journal} {Optica}\ }\textbf {\bibinfo {volume} {4}},\ \bibinfo {pages} {1298} (\bibinfo {year} {2017})}\BibitemShut {NoStop}%
\bibitem [{\citenamefont {Caizergues}\ \emph {et~al.}(2020)\citenamefont {Caizergues}, \citenamefont {Smartsev}, \citenamefont {Malka},\ and\ \citenamefont {Thaury}}]{caizergues2020phase}%
  \BibitemOpen
  \bibfield  {author} {\bibinfo {author} {\bibfnamefont {C.}~\bibnamefont {Caizergues}}, \bibinfo {author} {\bibfnamefont {S.}~\bibnamefont {Smartsev}}, \bibinfo {author} {\bibfnamefont {V.}~\bibnamefont {Malka}},\ and\ \bibinfo {author} {\bibfnamefont {C.}~\bibnamefont {Thaury}},\ }\bibfield  {title} {\bibinfo {title} {Phase-locked laser-wakefield electron acceleration},\ }\href {https://doi.org/10.1038/s41566-020-0657-2} {\bibfield  {journal} {\bibinfo  {journal} {Nature Photonics}\ }\textbf {\bibinfo {volume} {14}},\ \bibinfo {pages} {475} (\bibinfo {year} {2020})}\BibitemShut {NoStop}%
\bibitem [{\citenamefont {Oubrerie}\ \emph {et~al.}(2022)\citenamefont {Oubrerie}, \citenamefont {Andriyash}, \citenamefont {Lahaye}, \citenamefont {Smartsev}, \citenamefont {Malka},\ and\ \citenamefont {Thaury}}]{oubrerie2022axiparabola}%
  \BibitemOpen
  \bibfield  {author} {\bibinfo {author} {\bibfnamefont {K.}~\bibnamefont {Oubrerie}}, \bibinfo {author} {\bibfnamefont {I.~A.}\ \bibnamefont {Andriyash}}, \bibinfo {author} {\bibfnamefont {R.}~\bibnamefont {Lahaye}}, \bibinfo {author} {\bibfnamefont {S.}~\bibnamefont {Smartsev}}, \bibinfo {author} {\bibfnamefont {V.}~\bibnamefont {Malka}},\ and\ \bibinfo {author} {\bibfnamefont {C.}~\bibnamefont {Thaury}},\ }\bibfield  {title} {\bibinfo {title} {Axiparabola: a new tool for high-intensity optics},\ }\href {https://doi.org/10.1088/2040-8986/ac556a} {\bibfield  {journal} {\bibinfo  {journal} {Journal of Optics}\ }\textbf {\bibinfo {volume} {24}},\ \bibinfo {pages} {045503} (\bibinfo {year} {2022})}\BibitemShut {NoStop}%
\bibitem [{\citenamefont {Oz}\ \emph {et~al.}(2007)\citenamefont {Oz}, \citenamefont {Deng}, \citenamefont {Katsouleas}, \citenamefont {Muggli}, \citenamefont {Barnes}, \citenamefont {Blumenfeld}, \citenamefont {Decker}, \citenamefont {Emma}, \citenamefont {Hogan}, \citenamefont {Ischebeck} \emph {et~al.}}]{oz2007ionization}%
  \BibitemOpen
  \bibfield  {author} {\bibinfo {author} {\bibfnamefont {E.}~\bibnamefont {Oz}}, \bibinfo {author} {\bibfnamefont {S.}~\bibnamefont {Deng}}, \bibinfo {author} {\bibfnamefont {T.}~\bibnamefont {Katsouleas}}, \bibinfo {author} {\bibfnamefont {P.}~\bibnamefont {Muggli}}, \bibinfo {author} {\bibfnamefont {C.}~\bibnamefont {Barnes}}, \bibinfo {author} {\bibfnamefont {I.}~\bibnamefont {Blumenfeld}}, \bibinfo {author} {\bibfnamefont {F.}~\bibnamefont {Decker}}, \bibinfo {author} {\bibfnamefont {P.}~\bibnamefont {Emma}}, \bibinfo {author} {\bibfnamefont {M.}~\bibnamefont {Hogan}}, \bibinfo {author} {\bibfnamefont {R.}~\bibnamefont {Ischebeck}}, \emph {et~al.},\ }\bibfield  {title} {\bibinfo {title} {Ionization-induced electron trapping in ultrarelativistic plasma wakes},\ }\href {https://doi.org/10.1103/PhysRevLett.98.084801} {\bibfield  {journal} {\bibinfo  {journal} {Physical Review Letters}\ }\textbf {\bibinfo {volume} {98}},\ \bibinfo {pages} {084801} (\bibinfo {year} {2007})}\BibitemShut {NoStop}%
\bibitem [{\citenamefont {Pak}\ \emph {et~al.}(2010)\citenamefont {Pak}, \citenamefont {Marsh}, \citenamefont {Martins}, \citenamefont {Lu}, \citenamefont {Mori},\ and\ \citenamefont {Joshi}}]{pak2010injection}%
  \BibitemOpen
  \bibfield  {author} {\bibinfo {author} {\bibfnamefont {A.}~\bibnamefont {Pak}}, \bibinfo {author} {\bibfnamefont {K.}~\bibnamefont {Marsh}}, \bibinfo {author} {\bibfnamefont {S.}~\bibnamefont {Martins}}, \bibinfo {author} {\bibfnamefont {W.}~\bibnamefont {Lu}}, \bibinfo {author} {\bibfnamefont {W.}~\bibnamefont {Mori}},\ and\ \bibinfo {author} {\bibfnamefont {C.}~\bibnamefont {Joshi}},\ }\bibfield  {title} {\bibinfo {title} {Injection and trapping of tunnel-ionized electrons into laser-produced wakes},\ }\href {https://doi.org/10.1103/PhysRevLett.104.025003} {\bibfield  {journal} {\bibinfo  {journal} {Physical Review Letters}\ }\textbf {\bibinfo {volume} {104}},\ \bibinfo {pages} {025003} (\bibinfo {year} {2010})}\BibitemShut {NoStop}%
\bibitem [{\citenamefont {McGuffey}\ \emph {et~al.}(2010)\citenamefont {McGuffey}, \citenamefont {Thomas}, \citenamefont {Schumaker}, \citenamefont {Matsuoka}, \citenamefont {Chvykov}, \citenamefont {Dollar}, \citenamefont {Kalintchenko}, \citenamefont {Yanovsky}, \citenamefont {Maksimchuk}, \citenamefont {Krushelnick} \emph {et~al.}}]{mcguffey2010ionization}%
  \BibitemOpen
  \bibfield  {author} {\bibinfo {author} {\bibfnamefont {C.}~\bibnamefont {McGuffey}}, \bibinfo {author} {\bibfnamefont {A.}~\bibnamefont {Thomas}}, \bibinfo {author} {\bibfnamefont {W.}~\bibnamefont {Schumaker}}, \bibinfo {author} {\bibfnamefont {T.}~\bibnamefont {Matsuoka}}, \bibinfo {author} {\bibfnamefont {V.}~\bibnamefont {Chvykov}}, \bibinfo {author} {\bibfnamefont {F.}~\bibnamefont {Dollar}}, \bibinfo {author} {\bibfnamefont {G.}~\bibnamefont {Kalintchenko}}, \bibinfo {author} {\bibfnamefont {V.}~\bibnamefont {Yanovsky}}, \bibinfo {author} {\bibfnamefont {A.}~\bibnamefont {Maksimchuk}}, \bibinfo {author} {\bibfnamefont {K.}~\bibnamefont {Krushelnick}}, \emph {et~al.},\ }\bibfield  {title} {\bibinfo {title} {Ionization induced trapping in a laser wakefield accelerator},\ }\href {https://doi.org/10.1103/PhysRevLett.104.025004} {\bibfield  {journal} {\bibinfo  {journal} {Physical Review Letters}\ }\textbf {\bibinfo {volume} {104}},\ \bibinfo {pages} {025004} (\bibinfo {year} {2010})}\BibitemShut {NoStop}%
\bibitem [{\citenamefont {Xu}\ \emph {et~al.}(2014{\natexlab{b}})\citenamefont {Xu}, \citenamefont {Hua}, \citenamefont {Li}, \citenamefont {Zhang}, \citenamefont {Yan}, \citenamefont {Du}, \citenamefont {Huang}, \citenamefont {Chen}, \citenamefont {Tang}, \citenamefont {Lu}, \citenamefont {Yu}, \citenamefont {An}, \citenamefont {Joshi},\ and\ \citenamefont {Mori}}]{Xu2014PRL}%
  \BibitemOpen
  \bibfield  {author} {\bibinfo {author} {\bibfnamefont {X.~L.}\ \bibnamefont {Xu}}, \bibinfo {author} {\bibfnamefont {J.~F.}\ \bibnamefont {Hua}}, \bibinfo {author} {\bibfnamefont {F.}~\bibnamefont {Li}}, \bibinfo {author} {\bibfnamefont {C.~J.}\ \bibnamefont {Zhang}}, \bibinfo {author} {\bibfnamefont {L.~X.}\ \bibnamefont {Yan}}, \bibinfo {author} {\bibfnamefont {Y.~C.}\ \bibnamefont {Du}}, \bibinfo {author} {\bibfnamefont {W.~H.}\ \bibnamefont {Huang}}, \bibinfo {author} {\bibfnamefont {H.~B.}\ \bibnamefont {Chen}}, \bibinfo {author} {\bibfnamefont {C.~X.}\ \bibnamefont {Tang}}, \bibinfo {author} {\bibfnamefont {W.}~\bibnamefont {Lu}}, \bibinfo {author} {\bibfnamefont {P.}~\bibnamefont {Yu}}, \bibinfo {author} {\bibfnamefont {W.}~\bibnamefont {An}}, \bibinfo {author} {\bibfnamefont {C.}~\bibnamefont {Joshi}},\ and\ \bibinfo {author} {\bibfnamefont {W.~B.}\ \bibnamefont {Mori}},\ }\bibfield  {title} {\bibinfo {title} {Phase-space dynamics of ionization injection in plasma-based accelerators},\ }\href
  {https://doi.org/10.1103/PhysRevLett.112.035003} {\bibfield  {journal} {\bibinfo  {journal} {Phys. Rev. Lett.}\ }\textbf {\bibinfo {volume} {112}},\ \bibinfo {pages} {035003} (\bibinfo {year} {2014}{\natexlab{b}})}\BibitemShut {NoStop}%
\bibitem [{sup()}]{supplement}%
  \BibitemOpen
  \href@noop {} {}\bibinfo {note} {See Supplemental Material for additional simulation details and extended data.}\BibitemShut {Stop}%
\bibitem [{\citenamefont {Delahaye}\ \emph {et~al.}(2014)\citenamefont {Delahaye}, \citenamefont {Adli}, \citenamefont {Gessner}, \citenamefont {Hogan}, \citenamefont {Raubenheimer}, \citenamefont {An}, \citenamefont {Joshi},\ and\ \citenamefont {Mori}}]{adli2013beam}%
  \BibitemOpen
  \bibfield  {author} {\bibinfo {author} {\bibfnamefont {J.-P.}\ \bibnamefont {Delahaye}}, \bibinfo {author} {\bibfnamefont {E.}~\bibnamefont {Adli}}, \bibinfo {author} {\bibfnamefont {S.}~\bibnamefont {Gessner}}, \bibinfo {author} {\bibfnamefont {M.}~\bibnamefont {Hogan}}, \bibinfo {author} {\bibfnamefont {T.}~\bibnamefont {Raubenheimer}}, \bibinfo {author} {\bibfnamefont {W.}~\bibnamefont {An}}, \bibinfo {author} {\bibfnamefont {C.}~\bibnamefont {Joshi}},\ and\ \bibinfo {author} {\bibfnamefont {W.}~\bibnamefont {Mori}},\ }\bibfield  {title} {\bibinfo {title} {A beam driven plasma-wakefield linear collider from higgs factory to multi-tev},\ }in\ \href {https://accelconf.web.cern.ch/IPAC2014/papers/thpri015.pdf} {\emph {\bibinfo {booktitle} {{Proceedings of the 5th International Particle Accelerator Conference (IPAC'14)}}}}\ (\bibinfo {year} {2014})\ pp.\ \bibinfo {pages} {3791--3793}\BibitemShut {NoStop}%
\bibitem [{\citenamefont {Abada}\ \emph {et~al.}(2019)\citenamefont {Abada}, \citenamefont {Abbrescia}, \citenamefont {AbdusSalam}, \citenamefont {Abdyukhanov}, \citenamefont {Abelleira~Fernandez}, \citenamefont {Abramov}, \citenamefont {Aburaia}, \citenamefont {Acar}, \citenamefont {Adzic}, \citenamefont {Agrawal} \emph {et~al.}}]{fcc2019fcc}%
  \BibitemOpen
  \bibfield  {author} {\bibinfo {author} {\bibfnamefont {A.}~\bibnamefont {Abada}}, \bibinfo {author} {\bibfnamefont {M.}~\bibnamefont {Abbrescia}}, \bibinfo {author} {\bibfnamefont {S.}~\bibnamefont {AbdusSalam}}, \bibinfo {author} {\bibfnamefont {I.}~\bibnamefont {Abdyukhanov}}, \bibinfo {author} {\bibfnamefont {J.}~\bibnamefont {Abelleira~Fernandez}}, \bibinfo {author} {\bibfnamefont {A.}~\bibnamefont {Abramov}}, \bibinfo {author} {\bibfnamefont {M.}~\bibnamefont {Aburaia}}, \bibinfo {author} {\bibfnamefont {A.}~\bibnamefont {Acar}}, \bibinfo {author} {\bibfnamefont {P.}~\bibnamefont {Adzic}}, \bibinfo {author} {\bibfnamefont {P.}~\bibnamefont {Agrawal}}, \emph {et~al.},\ }\bibfield  {title} {\bibinfo {title} {Fcc-ee: The lepton collider: Future circular collider conceptual design report volume 2},\ }\href {https://doi.org/10.1140/epjst/e2019-900045-4} {\bibfield  {journal} {\bibinfo  {journal} {The European Physical Journal Special Topics}\ }\textbf {\bibinfo {volume} {228}},\ \bibinfo {pages} {261}
  (\bibinfo {year} {2019})}\BibitemShut {NoStop}%
\bibitem [{\citenamefont {Raubenheimer}\ \emph {et~al.}(2018)\citenamefont {Raubenheimer} \emph {et~al.}}]{raubenheimer2018lcls}%
  \BibitemOpen
  \bibfield  {author} {\bibinfo {author} {\bibfnamefont {T.}~\bibnamefont {Raubenheimer}} \emph {et~al.},\ }\bibfield  {title} {\bibinfo {title} {The lcls-ii-he, a high energy upgrade of the lcls-ii},\ }in\ \href {https://accelconf.web.cern.ch/fls2018/papers/mop1wd02.pdf} {\emph {\bibinfo {booktitle} {60th ICFA Advanced Beam Dynamics Workshop on Future Light Sources}}}\ (\bibinfo {year} {2018})\ pp.\ \bibinfo {pages} {6--11}\BibitemShut {NoStop}%
\bibitem [{\citenamefont {Schroeder}\ \emph {et~al.}(2014)\citenamefont {Schroeder}, \citenamefont {Vay}, \citenamefont {Esarey}, \citenamefont {Bulanov}, \citenamefont {Benedetti}, \citenamefont {Yu}, \citenamefont {Chen}, \citenamefont {Geddes},\ and\ \citenamefont {Leemans}}]{schroeder2014thermal}%
  \BibitemOpen
  \bibfield  {author} {\bibinfo {author} {\bibfnamefont {C.}~\bibnamefont {Schroeder}}, \bibinfo {author} {\bibfnamefont {J.-L.}\ \bibnamefont {Vay}}, \bibinfo {author} {\bibfnamefont {E.}~\bibnamefont {Esarey}}, \bibinfo {author} {\bibfnamefont {S.}~\bibnamefont {Bulanov}}, \bibinfo {author} {\bibfnamefont {C.}~\bibnamefont {Benedetti}}, \bibinfo {author} {\bibfnamefont {L.-L.}\ \bibnamefont {Yu}}, \bibinfo {author} {\bibfnamefont {M.}~\bibnamefont {Chen}}, \bibinfo {author} {\bibfnamefont {C.}~\bibnamefont {Geddes}},\ and\ \bibinfo {author} {\bibfnamefont {W.}~\bibnamefont {Leemans}},\ }\bibfield  {title} {\bibinfo {title} {Thermal emittance from ionization-induced trapping in plasma accelerators},\ }\href {https://doi.org/10.1103/PhysRevSTAB.17.101301} {\bibfield  {journal} {\bibinfo  {journal} {Physical Review Special Topics-Accelerators and Beams}\ }\textbf {\bibinfo {volume} {17}},\ \bibinfo {pages} {101301} (\bibinfo {year} {2014})}\BibitemShut {NoStop}%
\bibitem [{\citenamefont {Tzoufras}\ \emph {et~al.}(2009)\citenamefont {Tzoufras}, \citenamefont {Lu}, \citenamefont {Tsung}, \citenamefont {Huang}, \citenamefont {Mori}, \citenamefont {Katsouleas}, \citenamefont {Vieira}, \citenamefont {Fonseca},\ and\ \citenamefont {Silva}}]{tzoufras2009POP}%
  \BibitemOpen
  \bibfield  {author} {\bibinfo {author} {\bibfnamefont {M.}~\bibnamefont {Tzoufras}}, \bibinfo {author} {\bibfnamefont {W.}~\bibnamefont {Lu}}, \bibinfo {author} {\bibfnamefont {F.~S.}\ \bibnamefont {Tsung}}, \bibinfo {author} {\bibfnamefont {C.}~\bibnamefont {Huang}}, \bibinfo {author} {\bibfnamefont {W.~B.}\ \bibnamefont {Mori}}, \bibinfo {author} {\bibfnamefont {T.}~\bibnamefont {Katsouleas}}, \bibinfo {author} {\bibfnamefont {J.}~\bibnamefont {Vieira}}, \bibinfo {author} {\bibfnamefont {R.}~\bibnamefont {Fonseca}},\ and\ \bibinfo {author} {\bibfnamefont {L.}~\bibnamefont {Silva}},\ }\bibfield  {title} {\bibinfo {title} {Beam loading by electrons in nonlinear plasma wakes},\ }\href {https://doi.org/10.1063/1.3118628} {\bibfield  {journal} {\bibinfo  {journal} {Physics of Plasmas}\ }\textbf {\bibinfo {volume} {16}},\ \bibinfo {pages} {056705} (\bibinfo {year} {2009})}\BibitemShut {NoStop}%
\bibitem [{\citenamefont {Hogan}\ \emph {et~al.}(2010)\citenamefont {Hogan}, \citenamefont {Raubenheimer}, \citenamefont {Seryi}, \citenamefont {Muggli}, \citenamefont {Katsouleas}, \citenamefont {Huang}, \citenamefont {Lu}, \citenamefont {An}, \citenamefont {Marsh}, \citenamefont {Mori} \emph {et~al.}}]{hogan2010plasma}%
  \BibitemOpen
  \bibfield  {author} {\bibinfo {author} {\bibfnamefont {M.~J.}\ \bibnamefont {Hogan}}, \bibinfo {author} {\bibfnamefont {T.~O.}\ \bibnamefont {Raubenheimer}}, \bibinfo {author} {\bibfnamefont {A.}~\bibnamefont {Seryi}}, \bibinfo {author} {\bibfnamefont {P.}~\bibnamefont {Muggli}}, \bibinfo {author} {\bibfnamefont {T.}~\bibnamefont {Katsouleas}}, \bibinfo {author} {\bibfnamefont {C.}~\bibnamefont {Huang}}, \bibinfo {author} {\bibfnamefont {W.}~\bibnamefont {Lu}}, \bibinfo {author} {\bibfnamefont {W.}~\bibnamefont {An}}, \bibinfo {author} {\bibfnamefont {K.~A.}\ \bibnamefont {Marsh}}, \bibinfo {author} {\bibfnamefont {W.~B.}\ \bibnamefont {Mori}}, \emph {et~al.},\ }\bibfield  {title} {\bibinfo {title} {Plasma wakefield acceleration experiments at facet},\ }\href {https://doi.org/10.1088/1367-2630/12/5/055030} {\bibfield  {journal} {\bibinfo  {journal} {New Journal of Physics}\ }\textbf {\bibinfo {volume} {12}},\ \bibinfo {pages} {055030} (\bibinfo {year} {2010})}\BibitemShut {NoStop}%
\bibitem [{\citenamefont {Frederico}\ \emph {et~al.}(2011)\citenamefont {Frederico}, \citenamefont {Hogan}, \citenamefont {Nosochkov}, \citenamefont {Litos},\ and\ \citenamefont {Raubenheimer}}]{frederico2011facet}%
  \BibitemOpen
  \bibfield  {author} {\bibinfo {author} {\bibfnamefont {J.}~\bibnamefont {Frederico}}, \bibinfo {author} {\bibfnamefont {M.}~\bibnamefont {Hogan}}, \bibinfo {author} {\bibfnamefont {Y.}~\bibnamefont {Nosochkov}}, \bibinfo {author} {\bibfnamefont {M.}~\bibnamefont {Litos}},\ and\ \bibinfo {author} {\bibfnamefont {T.}~\bibnamefont {Raubenheimer}},\ }\href@noop {} {\emph {\bibinfo {title} {FACET Emittance Growth}}},\ \bibinfo {type} {Tech. Rep.}\ \bibinfo {number} {SLAC-PUB-14441}\ (\bibinfo  {institution} {SLAC National Accelerator Laboratory},\ \bibinfo {year} {2011})\BibitemShut {NoStop}%
\bibitem [{\citenamefont {Pogorelsky}\ \emph {et~al.}(2024)\citenamefont {Pogorelsky}, \citenamefont {Polyanskiy}, \citenamefont {Babzien}, \citenamefont {Simmonds},\ and\ \citenamefont {Palmer}}]{Pogorelsky2024}%
  \BibitemOpen
  \bibfield  {author} {\bibinfo {author} {\bibfnamefont {I.~V.}\ \bibnamefont {Pogorelsky}}, \bibinfo {author} {\bibfnamefont {M.~N.}\ \bibnamefont {Polyanskiy}}, \bibinfo {author} {\bibfnamefont {M.}~\bibnamefont {Babzien}}, \bibinfo {author} {\bibfnamefont {A.}~\bibnamefont {Simmonds}},\ and\ \bibinfo {author} {\bibfnamefont {M.~A.}\ \bibnamefont {Palmer}},\ }\bibfield  {title} {\bibinfo {title} {Terawatt-class femtosecond long-wave infrared laser},\ }\href {https://doi.org/10.3389/fphy.2024.1390225} {\bibfield  {journal} {\bibinfo  {journal} {Frontiers in Physics}\ }\textbf {\bibinfo {volume} {12}},\ \bibinfo {pages} {1390225} (\bibinfo {year} {2024})}\BibitemShut {NoStop}%
\bibitem [{\citenamefont {Fonseca}\ \emph {et~al.}(2002)\citenamefont {Fonseca}, \citenamefont {Silva}, \citenamefont {Tsung}, \citenamefont {Decyk}, \citenamefont {Lu}, \citenamefont {Ren}, \citenamefont {Mori}, \citenamefont {Deng}, \citenamefont {Lee}, \citenamefont {Katsouleas},\ and\ \citenamefont {Adam}}]{Fonseca2002}%
  \BibitemOpen
  \bibfield  {author} {\bibinfo {author} {\bibfnamefont {R.~A.}\ \bibnamefont {Fonseca}}, \bibinfo {author} {\bibfnamefont {L.~O.}\ \bibnamefont {Silva}}, \bibinfo {author} {\bibfnamefont {F.~S.}\ \bibnamefont {Tsung}}, \bibinfo {author} {\bibfnamefont {V.~K.}\ \bibnamefont {Decyk}}, \bibinfo {author} {\bibfnamefont {W.}~\bibnamefont {Lu}}, \bibinfo {author} {\bibfnamefont {C.}~\bibnamefont {Ren}}, \bibinfo {author} {\bibfnamefont {W.~B.}\ \bibnamefont {Mori}}, \bibinfo {author} {\bibfnamefont {S.}~\bibnamefont {Deng}}, \bibinfo {author} {\bibfnamefont {S.}~\bibnamefont {Lee}}, \bibinfo {author} {\bibfnamefont {T.}~\bibnamefont {Katsouleas}},\ and\ \bibinfo {author} {\bibfnamefont {J.~C.}\ \bibnamefont {Adam}},\ }\bibfield  {title} {\bibinfo {title} {Osiris: A three-dimensional, fully relativistic particle in cell code for modeling plasma based accelerators},\ }in\ \href {https://doi.org/10.1007/3-540-47789-6_36} {\emph {\bibinfo {booktitle} {Computational Science --- ICCS 2002}}}\ (\bibinfo  {publisher}
  {Springer Berlin Heidelberg},\ \bibinfo {address} {Berlin, Heidelberg},\ \bibinfo {year} {2002})\ pp.\ \bibinfo {pages} {342--351}\BibitemShut {NoStop}%
\bibitem [{\citenamefont {Davidson}\ \emph {et~al.}(2015)\citenamefont {Davidson}, \citenamefont {Tableman}, \citenamefont {An}, \citenamefont {Tsung}, \citenamefont {Lu}, \citenamefont {Vieira}, \citenamefont {Fonseca}, \citenamefont {Silva},\ and\ \citenamefont {Mori}}]{Davidson2015}%
  \BibitemOpen
  \bibfield  {author} {\bibinfo {author} {\bibfnamefont {A.}~\bibnamefont {Davidson}}, \bibinfo {author} {\bibfnamefont {A.}~\bibnamefont {Tableman}}, \bibinfo {author} {\bibfnamefont {W.}~\bibnamefont {An}}, \bibinfo {author} {\bibfnamefont {F.}~\bibnamefont {Tsung}}, \bibinfo {author} {\bibfnamefont {W.}~\bibnamefont {Lu}}, \bibinfo {author} {\bibfnamefont {J.}~\bibnamefont {Vieira}}, \bibinfo {author} {\bibfnamefont {R.}~\bibnamefont {Fonseca}}, \bibinfo {author} {\bibfnamefont {L.}~\bibnamefont {Silva}},\ and\ \bibinfo {author} {\bibfnamefont {W.}~\bibnamefont {Mori}},\ }\bibfield  {title} {\bibinfo {title} {Implementation of a hybrid particle code with a {PIC} description in r{\textendash}z and a gridless description in $\phi$ into {OSIRIS}},\ }\href {https://doi.org/10.1016/j.jcp.2014.10.064} {\bibfield  {journal} {\bibinfo  {journal} {Journal of Computational Physics}\ }\textbf {\bibinfo {volume} {281}},\ \bibinfo {pages} {1063} (\bibinfo {year} {2015})}\BibitemShut {NoStop}%
\bibitem [{\citenamefont {Li}\ \emph {et~al.}(2021{\natexlab{a}})\citenamefont {Li}, \citenamefont {Miller}, \citenamefont {Xu}, \citenamefont {Tsung}, \citenamefont {Decyk}, \citenamefont {An}, \citenamefont {Fonseca},\ and\ \citenamefont {Mori}}]{LI2021107580}%
  \BibitemOpen
  \bibfield  {author} {\bibinfo {author} {\bibfnamefont {F.}~\bibnamefont {Li}}, \bibinfo {author} {\bibfnamefont {K.~G.}\ \bibnamefont {Miller}}, \bibinfo {author} {\bibfnamefont {X.}~\bibnamefont {Xu}}, \bibinfo {author} {\bibfnamefont {F.~S.}\ \bibnamefont {Tsung}}, \bibinfo {author} {\bibfnamefont {V.~K.}\ \bibnamefont {Decyk}}, \bibinfo {author} {\bibfnamefont {W.}~\bibnamefont {An}}, \bibinfo {author} {\bibfnamefont {R.~A.}\ \bibnamefont {Fonseca}},\ and\ \bibinfo {author} {\bibfnamefont {W.~B.}\ \bibnamefont {Mori}},\ }\bibfield  {title} {\bibinfo {title} {A new field solver for modeling of relativistic particle-laser interactions using the particle-in-cell algorithm},\ }\href {https://doi.org/10.1016/j.cpc.2020.107580} {\bibfield  {journal} {\bibinfo  {journal} {Computer Physics Communications}\ }\textbf {\bibinfo {volume} {258}},\ \bibinfo {pages} {107580} (\bibinfo {year} {2021}{\natexlab{a}})}\BibitemShut {NoStop}%
\bibitem [{\citenamefont {Bruhwiler}\ \emph {et~al.}(2003)\citenamefont {Bruhwiler}, \citenamefont {Dimitrov}, \citenamefont {Cary}, \citenamefont {Esarey}, \citenamefont {Leemans},\ and\ \citenamefont {Giacone}}]{bruhwiler2003particle}%
  \BibitemOpen
  \bibfield  {author} {\bibinfo {author} {\bibfnamefont {D.~L.}\ \bibnamefont {Bruhwiler}}, \bibinfo {author} {\bibfnamefont {D.}~\bibnamefont {Dimitrov}}, \bibinfo {author} {\bibfnamefont {J.~R.}\ \bibnamefont {Cary}}, \bibinfo {author} {\bibfnamefont {E.}~\bibnamefont {Esarey}}, \bibinfo {author} {\bibfnamefont {W.}~\bibnamefont {Leemans}},\ and\ \bibinfo {author} {\bibfnamefont {R.~E.}\ \bibnamefont {Giacone}},\ }\bibfield  {title} {\bibinfo {title} {Particle-in-cell simulations of tunneling ionization effects in plasma-based accelerators},\ }\href {https://doi.org/10.1063/1.1566935} {\bibfield  {journal} {\bibinfo  {journal} {Physics of Plasmas}\ }\textbf {\bibinfo {volume} {10}},\ \bibinfo {pages} {2022} (\bibinfo {year} {2003})}\BibitemShut {NoStop}%
\bibitem [{\citenamefont {Li}\ \emph {et~al.}(2021{\natexlab{b}})\citenamefont {Li}, \citenamefont {An}, \citenamefont {Decyk}, \citenamefont {Xu}, \citenamefont {Hogan},\ and\ \citenamefont {Mori}}]{LI2021107784}%
  \BibitemOpen
  \bibfield  {author} {\bibinfo {author} {\bibfnamefont {F.}~\bibnamefont {Li}}, \bibinfo {author} {\bibfnamefont {W.}~\bibnamefont {An}}, \bibinfo {author} {\bibfnamefont {V.~K.}\ \bibnamefont {Decyk}}, \bibinfo {author} {\bibfnamefont {X.}~\bibnamefont {Xu}}, \bibinfo {author} {\bibfnamefont {M.~J.}\ \bibnamefont {Hogan}},\ and\ \bibinfo {author} {\bibfnamefont {W.~B.}\ \bibnamefont {Mori}},\ }\bibfield  {title} {\bibinfo {title} {A quasi-static particle-in-cell algorithm based on an azimuthal fourier decomposition for highly efficient simulations of plasma-based acceleration: Qpad},\ }\href {https://doi.org/10.1016/j.cpc.2020.107784} {\bibfield  {journal} {\bibinfo  {journal} {Computer Physics Communications}\ }\textbf {\bibinfo {volume} {261}},\ \bibinfo {pages} {107784} (\bibinfo {year} {2021}{\natexlab{b}})}\BibitemShut {NoStop}%
\bibitem [{\citenamefont {Pierce}\ \emph {et~al.}(2023)\citenamefont {Pierce}, \citenamefont {Palastro}, \citenamefont {Li}, \citenamefont {Malaca}, \citenamefont {Ramsey}, \citenamefont {Vieira}, \citenamefont {Weichman},\ and\ \citenamefont {Mori}}]{pierce2023arbitrarily}%
  \BibitemOpen
  \bibfield  {author} {\bibinfo {author} {\bibfnamefont {J.~R.}\ \bibnamefont {Pierce}}, \bibinfo {author} {\bibfnamefont {J.~P.}\ \bibnamefont {Palastro}}, \bibinfo {author} {\bibfnamefont {F.}~\bibnamefont {Li}}, \bibinfo {author} {\bibfnamefont {B.}~\bibnamefont {Malaca}}, \bibinfo {author} {\bibfnamefont {D.}~\bibnamefont {Ramsey}}, \bibinfo {author} {\bibfnamefont {J.}~\bibnamefont {Vieira}}, \bibinfo {author} {\bibfnamefont {K.}~\bibnamefont {Weichman}},\ and\ \bibinfo {author} {\bibfnamefont {W.~B.}\ \bibnamefont {Mori}},\ }\bibfield  {title} {\bibinfo {title} {Arbitrarily structured laser pulses},\ }\href {https://doi.org/10.1103/PhysRevResearch.5.013085} {\bibfield  {journal} {\bibinfo  {journal} {Physical Review Research}\ }\textbf {\bibinfo {volume} {5}},\ \bibinfo {pages} {013085} (\bibinfo {year} {2023})}\BibitemShut {NoStop}%
\bibitem [{\citenamefont {Franke}\ \emph {et~al.}(2021)\citenamefont {Franke}, \citenamefont {Ramsey}, \citenamefont {Simpson}, \citenamefont {Turnbull}, \citenamefont {Froula},\ and\ \citenamefont {Palastro}}]{franke2021optical}%
  \BibitemOpen
  \bibfield  {author} {\bibinfo {author} {\bibfnamefont {P.}~\bibnamefont {Franke}}, \bibinfo {author} {\bibfnamefont {D.}~\bibnamefont {Ramsey}}, \bibinfo {author} {\bibfnamefont {T.~T.}\ \bibnamefont {Simpson}}, \bibinfo {author} {\bibfnamefont {D.}~\bibnamefont {Turnbull}}, \bibinfo {author} {\bibfnamefont {D.}~\bibnamefont {Froula}},\ and\ \bibinfo {author} {\bibfnamefont {J.}~\bibnamefont {Palastro}},\ }\bibfield  {title} {\bibinfo {title} {Optical shock-enhanced self-photon acceleration},\ }\href {https://doi.org/10.1103/PhysRevA.104.043520} {\bibfield  {journal} {\bibinfo  {journal} {Physical Review A}\ }\textbf {\bibinfo {volume} {104}},\ \bibinfo {pages} {043520} (\bibinfo {year} {2021})}\BibitemShut {NoStop}%
\bibitem [{\citenamefont {Xu}(2020)}]{xu2020phase}%
  \BibitemOpen
  \bibfield  {author} {\bibinfo {author} {\bibfnamefont {X.}~\bibnamefont {Xu}},\ }\href@noop {} {\emph {\bibinfo {title} {Phase Space Dynamics in Plasma Based Wakefield Acceleration}}},\ Springer Theses\ (\bibinfo  {publisher} {Springer Nature Singapore},\ \bibinfo {address} {Singapore},\ \bibinfo {year} {2020})\BibitemShut {NoStop}%
\end{thebibliography}
\end{document}


\renewcommand{\thetable}{S\arabic{table}}
\renewcommand{\thefigure}{S\arabic{figure}}
\renewcommand{\thesection}{S\arabic{section}}

\title{Supplementary Information:\\
\normalsize{Collider-quality electron bunches from an all-optical plasma photoinjector}}

\author{Arohi Jain}
\affiliation{Department of Physics and Astronomy, Stony Brook University, Stony Brook, USA}
\author{Jiayang Yan}
\affiliation{Department of Physics and Astronomy, Stony Brook University, Stony Brook, USA}
\author{Jacob R. Pierce}
\affiliation{ Department of Electrical and Computer Engineering, University of California, Los Angeles, California 90095, USA}

\author{Tanner T. Simpson}
\affiliation{University of Rochester, Laboratory for Laser Energetics, Rochester, New York 14623, USA}

\author{Mikhail Polyanskiy}
\affiliation{Accelerator Test Facility, Brookhaven National Laboratory, Upton, NY 11973, USA}
\author{William Li}
\affiliation{Accelerator Test Facility, Brookhaven National Laboratory, Upton, NY 11973, USA}
\author{Marcus Babzien}
\affiliation{Accelerator Test Facility, Brookhaven National Laboratory, Upton, NY 11973, USA}
\author{Mark Palmer}
\affiliation{Accelerator Test Facility, Brookhaven National Laboratory, Upton, NY 11973, USA}

\author{Michael Downer}
\affiliation{Department of Physics, University of Texas at Austin, Austin TX 78712, USA}
\author{Roman Samulyak}
\affiliation{Department of Physics and Astronomy, Stony Brook University, Stony Brook, USA}
\author{Chan Joshi}
\affiliation{ Department of Electrical and Computer Engineering, University of California, Los Angeles, California 90095, USA}
\author{Warren B. Mori}
\affiliation{ Department of Electrical and Computer Engineering, University of California, Los Angeles, California 90095, USA}

\author{John P. Palastro}
\affiliation{University of Rochester, Laboratory for Laser Energetics, Rochester, New York 14623, USA}
\author{Navid Vafaei-Najafabadi}
\affiliation{Department of Physics and Astronomy, Stony Brook University, Stony Brook, USA}

\maketitle
\section{High-charge injection using Conventional pulse }\label{high-charge}

The conventional injector pulse produces a triangular-shaped electron bunch with 236 pC of charge using a 15 $\mu$m spot size [Fig. S1]. However, the resulting emittance ($\epsilon_x = 1.3$ $\mu$m-rad, $\epsilon_y = 0.69$ $\mu$m-rad) is too large to meet the Snowmass requirements, making it unsuitable for collider applications. Additionally, this triangular profile fails to flatten the longitudinal field, leading to a suboptimal energy spread in the acceleration stage. In contrast, the flying focus pulse injects the similar charge with trapezoidal current profile [Fig. 2 of the main manuscript] using a significantly smaller 8.6 $\mu$m spot size, demonstrating improved beam quality. The reduced emittance and optimized charge distribution in the flying focus scheme ensure compatibility with collider requirements, making it a superior alternative to conventional injection methods.

\begin{figure*}[!t]
\centering
\includegraphics[width=0.6\textwidth]{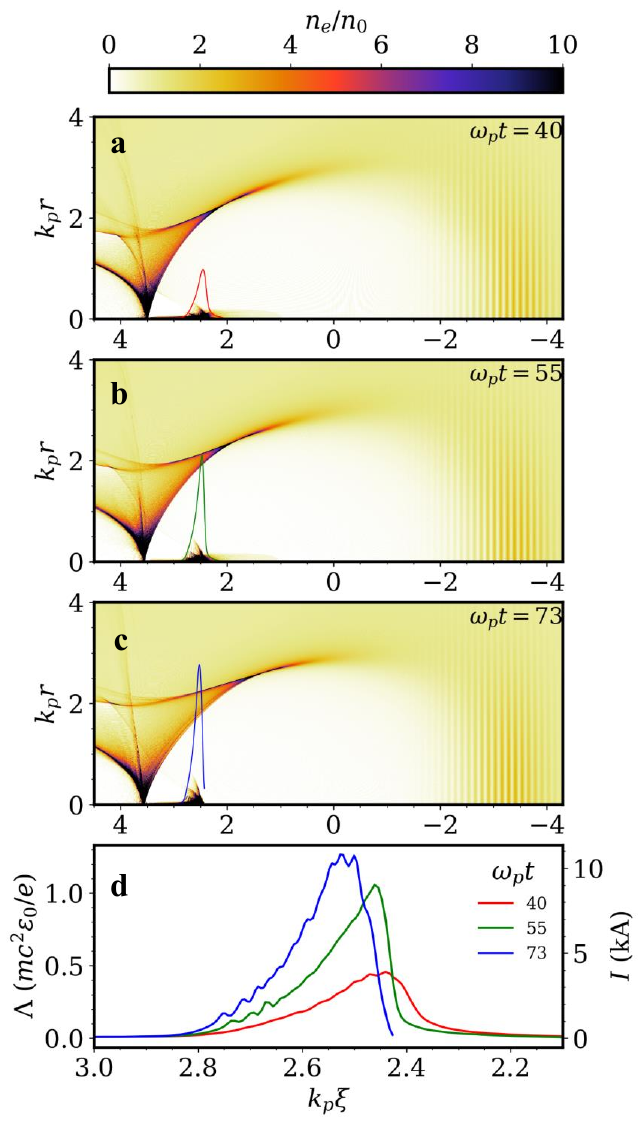}
\vfil
\caption{Ionization injection using a conventional laser pulse.Simulations of ionization injection into a CO$_2$ laser-driven wake are shown in (a-c), where a conventional laser pulse serves as the injector. Both laser pulses propagate from left to right. The contour represents the electron density ($n_e$).(d) Linear charge density ($\Lambda$) and current ($I$) are shown at different times. The final triangular bunch carries a charge of 236 pC, but simulations indicate that the resulting emittance ($\epsilon_x = 1.3$ $\mathrm{\mu}$m rad, $\epsilon_y = 0.69$ $\mu$m rad) is too large to meet Snowmass requirements \cite{benedetti2022whitepaper}. Laser pulse parameters: CO$_2$ driver laser pulse – 5.3 J, 250 fs FWHM, 9.2 $\mu$m wavelength; conventional laser pulse – 0.15 J, 160 fs FWHM, 15 $\mu$m spot size, 0.4 $\mu$m wavelength.}
\label{fig:conv}
\end{figure*}

\begin{figure*}[!t]
\centering
\includegraphics[width=\textwidth]{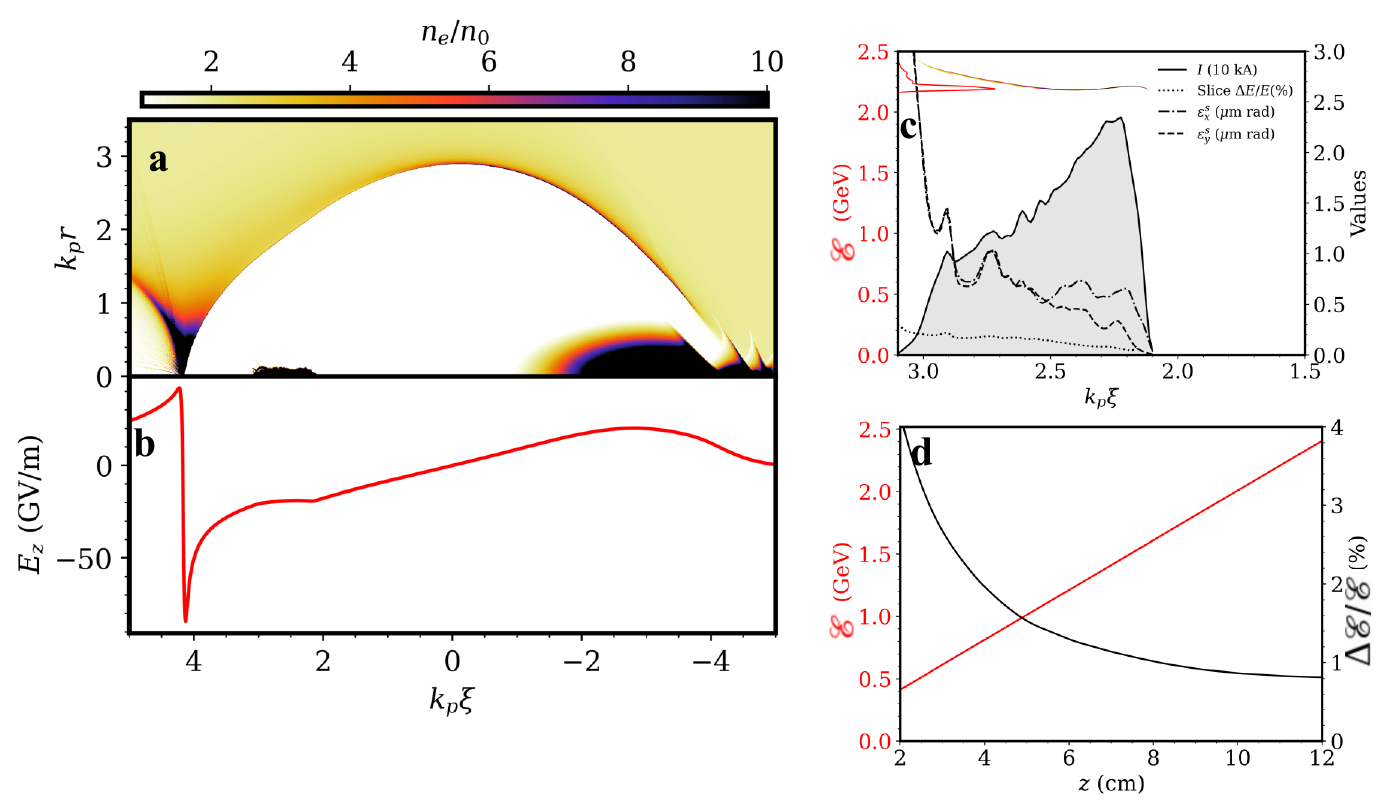}
\vfil
\caption{Acceleration of trapezoidal nC electron bunch in PWFA. (a) Generated trapezoidal electron bunch inside the wake driven by electron beam driver. \textbf{b} Witness shaped electron bunch flattens the wakefield. (c) Witness bunch longitudinal phase space (left y-axis) and slice current $I$, slice relative energy spread $\Delta \mathscr{E/ E}$ and transverse slice emittances $\epsilon_n^s$ (right y-axis). (d) Evolution of energy (left y-axis) and FWHM energy spread (right y-axis) with propagation distance $z$.}
\label{fig:nC beam}
\end{figure*}

\section{Acceleration of nano-Coulomb bunch}
Operating laser-plasma accelerators in the high-current regime underscores their potential as drivers for next-generation compact light sources \cite{galletti2024prospects}. The flying focus injection scheme presented in the main text enables the generation of nanocoulomb-class electron bunches. In this configuration, a flying focus pulse with a spot size of $w_0 = 16.5 , \mu\mathrm{m}$ is used. The effective ionization length is given by $L_i = L + 2\sqrt{3} \sqrt{\ln(a_0/a_T)} Z_R$, where $L = \Delta \xi_i /(1/v_F - 1)$ is the focal range. This setup, along with the parameters listed in Table I of the main manuscript, yields the desired high-charge bunch.

\vspace{6pt}
The acceleration stage is implemented using the same methodology as described in the main paper. A beam driver with a charge of $Q = 3.2 \ \mathrm{nC}$ is considered, with other beam parameters listed in Table II of the main manuscript. These settings are consistent with those of electron beams previously available at the Facility for Advanced Accelerator Experimental Tests (FACET-I) at SLAC National Accelerator Laboratory \cite{hogan2011plasma}. The injected bunch effectively beamloads the wakefield, as shown in Fig.\ref{fig:nC beam}b. While the energy phase space remains flat in the high-current region, the rear of the bunch develops higher emittance due to space charge effects, as shown in Fig.\ref{fig:nC beam}c. Figure~\ref{fig:nC beam}d demonstrates that the nanocoulomb-class trapezoidal bunch is efficiently accelerated over $10 \ \mathrm{cm}$ to approximately $2 \ \mathrm{GeV}$, maintaining an energy spread below $1\%$.

\section{Chromatic Flying Focus}\label{Chromatic flying focus}

The flying focus is an optical technique that enables spatiotemporal control over the trajectory of a focal point, allowing it to traverse distances far beyond the Rayleigh range while remaining independent of the laser's group velocity \cite{sainte2017controlling, froula2018spatiotemporal}. This was first demonstrated experimentally using chromatic focusing of a chirped laser pulse \cite{froula2018spatiotemporal}. Other approaches to achieve flying focus include achromatic flying focus methods based on space-time light sheets \cite{yessenov2020accelerating, kondakci2017diffraction}, axiparabola-echelon mirrors \cite{Palastro2020dephasing}, and nonlinear media \cite{simpson2020nonlinear, simpson2022spatiotemporal}.  In this study, a chromatic flying focus pulse serves as the injector laser in a two-color ionization scheme.

\vspace{6pt}
The injector pulse is linearly chirped, with the chirp value determining the velocity of the flying focus. The spot size is set at $8.6~\mu m$, optimizing the emittance to be in the tens of nanometers. The focal range \( L \) is selected to produce a 200 pC charge. The velocity of the flying focus \( {v}_F = 1.01 \) is determined by the equation \( \Delta \xi_i = \left( 1/{v}_F - 1 \right) L \), while the longitudinal focal shift \( L \) is given by \( L = f_0 \Delta \lambda / \lambda \), where \( \Delta \lambda \) is the laser bandwidth, \( \lambda \) is the central wavelength, and \( f_0 \) is the focal length. The incident pulse duration \( T \) is related to the flying focus velocity \cite{froula2019flying} by \( v_F = c(1 \pm T / L c)^{-1} \). The transform-limited pulse duration \( \tau \) is approximately \( \tau \approx T / |\eta| \), which also corresponds to the ionization pulse duration of the conventional laser pulse in Fig. 2 of the manuscript. The velocity \( v_f \) is connected to the chirp as \( v_F = c / (1 + \eta c \tau^2 w_0 / (2 f_0)) \) \cite{palastro2018ionization}, where \( \eta \) is the chirp parameter, \( \tau \) is the transform-limited pulse duration, \( w_0 \) is the initial spot size, and \( f_0 \) is the nominal focal length. Using an incident laser with \( f_0 = 0.15 \) m, \( \tau = 20 \) fs, and chirp = 7.6, a focal range \( L \) of 3.8 mm is achieved, providing the required flying focus pulse parameters for the experiment. The location of the focus spot (or ionization point) \( \xi_i \), in the co-moving coordinate system, is given by:
\begin{equation}
    \xi_i = \xi_{i0} + \frac{\eta c \tau^2 \omega_0}{2} \left( \frac{z}{f_0} -1 \right), \label{eqn: location of focus xi}
\end{equation}

\section{Experimental Preparation and Feasibility}

The experimental realization of this work at Accelerator Test Facility (ATF), Brookhaven National Laboratory depends on mitigating key technical challenges. The primary challenge, producing a sub-picosecond CO$_2$ laser pulse, is addressed in the main manuscript. Furthermore, the facility’s unique laser sources, combined with its development of high-precision, multi-beam platforms, enable inherently robust alignment and synchronization essential for the study presented in this paper. The following points articulate the key capabilities and design advantages that ensure the experiment's feasibility.

\begin{enumerate}
    \item \textbf{Large plasma bubble and relaxed synchronization requirements.} \\
    In the 220~pC configuration presented in the paper, the CO$_2$-driven bubble has a diameter of approximately 200~$\mu$m. The ability of the CO$_2$ laser to generate macroscopic plasma bubbles is a key characteristic of this laser system and serves as an enabling advantage for the present experiment. The larger scale of the structure greatly relaxes the alignment constraints for two-color ionization injection, making them far less stringent than the femtosecond synchronization requirements typically needed when using an NIR ($\lambda \sim 1~\mu$m) driver. We provide the expected experimental tolerances below, but it is important to note that the experiment is inherently tolerant to moderate jitter. Even at a 100~fs ($\sim$30~$\mu$m) jitter—corresponding to roughly 30\% of the plasma-bubble size—the flying focus pulse stays well within the bubble, resulting mainly in a change in the characteristics of the injected trapezoid beam rather than compromising injection. To directly address this, we ran a sample simulation applying a 100~fs timing jitter to the flying focus pulse. The results show that the bunch is still successfully captured and accelerated. While the ionization occurs at a slightly different phase, resulting in a modified trapezoidal shape compared to the case with zero jitter, both the overall trapezoidal profile and the low emittance are preserved.  Specifically, the bunch subjected to the 100~fs jitter carries a charge of 212 pC and maintains ultra-low emittances of $\epsilon_x = 180$ nm rad and $\epsilon_y = 78$ nm rad. This is highly comparable to the ideal zero-jitter case, which yields a 220 pC charge and emittances of $\epsilon_x = 171$ nm rad and $\epsilon_y = 76$ nm rad. This confirms that our scheme continues to work and produce high-quality bunches even under realistic jitter conditions.

    \item \textbf{Pointing stability of the Ti:Sapphire laser.} \\
    The pointing stability of the Ti:Sapphire laser at focus has been measured to $<$1~$\mu$m position jitter for a 4~$\mu$m focal spot, corresponding to 25~$\mu$rad STD~\cite{kupfer2020}. For the 8~$\mu$m focal spot used in the 220~pC scenario discussed in the paper, the pointing stability is therefore expected to be $\sim$2~$\mu$m.

    \item \textbf{Timing system upgrades.} \\
    The ATF upgrade roadmap includes the transition of the facility’s timing distribution system to an optical platform, as part of ongoing modernization efforts to support precision multi-beam experiments. As a benchmark for achievable performance, state-of-the-art optical timing systems provide specifications of RMS jitter near 5~fs ($\sim$1.5~$\mu$m)~\cite{CyclePULSE}.

    \item \textbf{Diagnostic and alignment capabilities.} \\
    In addition to the unique combination of CO$_2$ and NIR laser sources, ATF provides a linac-produced electron beam, which has been used experimentally for the transverse electron radiography of CO$_2$-generated plasma structures in density ranges of $3 \times 10^{15}~\text{cm}^{-3}$ to $1 \times 10^{17}~\text{cm}^{-3}$~\cite{VafaeiNajafabadi2024,wu2024}. The sensitivity of the electron radiography to the fields of the laser and plasma serves as a key enabling technology for diagnosing laser–plasma interactions, including the alignment and synchronization between the two lasers \textit{in situ}.
\end{enumerate}

Collectively, these existing and planned capabilities at ATF provide strong assurance of the feasibility and success of the proposed experiment. These details are provided here to clarify the experimental readiness and robustness of the setup supporting this work.